\RequirePackage{ifpdf}
\ifpdf % We are running pdfTeX in pdf mode
\documentclass[pdftex]{sigma}
\else
\documentclass{sigma}
\fi

\numberwithin{equation}{section}

\begin{document}

\allowdisplaybreaks

\renewcommand{\thefootnote}{$\star$}

\renewcommand{\PaperNumber}{007}

\FirstPageHeading

\ShortArticleName{On the Spectrum of a Discrete Non-Hermitian Quantum System}

\ArticleName{On the Spectrum of a Discrete\\ Non-Hermitian Quantum System\footnote{This paper is a contribution to the Proceedings of the VIIth Workshop ``Quantum Physics with Non-Hermitian Operators''
     (June 29 -- July 11, 2008, Benasque, Spain). The full collection
is available at
\href{http://www.emis.de/journals/SIGMA/PHHQP2008.html}{http://www.emis.de/journals/SIGMA/PHHQP2008.html}}}

\Author{Ebru ERGUN}

\AuthorNameForHeading{E. Ergun}

\Address{Department of Physics, Ankara University, 06100 Tandogan, Ankara, Turkey} % Address of First Author
\Email{\href{mailto:eergun@science.ankara.edu.tr}{eergun@science.ankara.edu.tr}}

\ArticleDates{Received October 28, 2008, in f\/inal form January 13,
2009; Published online January 19, 2009}

\Abstract{In this paper, we develop spectral analysis of a discrete non-Hermitian quantum system that is a discrete counterpart of some continuous quantum systems on a complex contour. In particular, simple conditions for
discreteness of the spectrum are established.}


\Classification{39A70; 81Q10}

\section{Introduction}\label{section1}

Non-Hermitian Hamiltonians and complex extension of quantum mechanics have
recently received a lot of attention (see review \cite{[1]}). This f\/ield of
mathematical physics came into f\/lower due to the eigenvalue problem%
\begin{equation}
-\psi ^{\prime \prime }(x)+ix^{3}\psi (x)=E\psi (x),\qquad -\infty
<x<\infty ,  \label{1.1}
\end{equation}%
where $\psi (x)$ is a desired solution, $i=\sqrt{-1}$ is the imaginary unit,
$E$ is a complex parameter (``energy'' or spectral parameter).

A complex value $E_{0}$ of the parameter $E$ is called an \textit{eigenvalue}
of equation (\ref{1.1}) if this equation with $E=E_{0}$ has a nontrivial
(non-identically zero) solution $\psi _{0}(x)$ that belongs to the Hilbert
space $L^{2}(-\infty ,\infty )$ of complex-valued functions $f$ def\/ined on $%
(-\infty ,\infty )$ such that%
\[
\int_{-\infty }^{\infty }\left\vert f(x)\right\vert ^{2}dx<\infty
\]%
with the inner (scalar) product%
\begin{equation}
\left\langle f,g\right\rangle =\int_{-\infty }^{\infty }f(x)\overline{g}(x)dx
\label{1.2}
\end{equation}%
in which the bar over a function denotes the complex conjugate. The function
$\psi _{0}(x)$ is called an \textit{eigenfunction} of equation (\ref{1.1}),
corresponding to the eigenvalue $E_{0}.$

If we def\/ine the operator $S:D\subset L^{2}(-\infty ,\infty )\rightarrow
L^{2}(-\infty ,\infty )$ with the domain $D$ consisting of the functions $%
f\in L^{2}(-\infty ,\infty )$ that are dif\/ferentiable, with the derivative $%
f^{\prime }$ absolutely continuous on each f\/inite subinterval of $(-\infty
,\infty )$ and such that%
\[
-f^{\prime \prime }+ix^{3}f\in L^{2}(-\infty ,\infty ),
\]%
by letting%
\begin{equation}
Sf=-f^{\prime \prime }+ix^{3}f\qquad \text{for} \quad f\in D,  \label{1.3}
\end{equation}%
then eigenvalue problem (\ref{1.1}) can be written as $S\psi =E\psi $, $\psi
\in D,$ $\psi \neq 0.$ Consequently, eigenvalues of equation (\ref{1.1}) are
eigenvalues of the operator~$S$.

Because of the fact that the operator $S$ def\/ined by (\ref{1.3}) has the
complex non-real coef\/f\/icient (potential) equal to $ix^{3},$ this operator is
not Hermitian with respect to the inner product (\ref{1.2}), that is, we
cannot state that%
\[
\left\langle Sf,g\right\rangle =\left\langle f,Sg\right\rangle \qquad \text{for all} \ \ f,g\in D.
\]%
Therefore it is not obvious that the eigenvalues of $S$ may be real
(remember that the eigenvalues of any Hermitian operator are real and the
eigenvectors corresponding to the distinct eigenvalues are orthogonal).

Around 1992 Bessis and Zinn-Justin had noticed on the basis of numerical
work that some of the eigenvalues of equation (\ref{1.1}) seemed to be real
and positive and they conjectured (not in print) that for equation (\ref{1.1}) the eigenvalues are all real and positive.

In 1998, Bender and Boettcher \cite{[2]} generalized the BZJ conjecture,
namely, they conjectured (again on the basis of numerical analysis) that the
eigenvalues of the equation%
\begin{equation}
-\psi ^{\prime \prime }(x)-(ix)^{m}\psi (x)=E\psi (x),\qquad -\infty
<x<\infty ,  \label{1.4}
\end{equation}%
are all real and positive provided $m\geq 2.$ Note that in equation (\ref{1.4}) $m$ is an arbitrary positive real number and equation (\ref{1.1})
corresponds to the choice $m=3$ in (\ref{1.4}).

Bender and Boettcher assumed that the reason for reality of eigenvalues of (%
\ref{1.4}) (in particular of (\ref{1.1})) must be certain symmetry property
of this equation, namely, the so-called $PT$-symmetry of it (for more
details see \cite{[1]}).

$P$ (parity) and $T$ (time reversal) operations are def\/ined by%
\[
Pf(x)=f(-x)\qquad \text{and} \qquad Tf(x)=\overline{f}(x),
\]%
respectively. A Hamiltonian%
\begin{equation}
H=-\frac{d^{2}}{dx^{2}}+V(x)  \label{1.5}
\end{equation}%
with complex potential $V(x)$ is called $PT$-symmetric if it commutes with
the composite operation~$PT$:%
\[
\left[ H,PT\right] =HPT-PTH=0.
\]%
It is easily seen that $PT$-symmetricity of $H$ given by (\ref{1.5}) is
equivalent to the condition%
\[
\overline{V}(-x)=V(x).
\]%
The potentials $ix^{3}$ and $-(ix)^{m}$ of equations (\ref{1.1}) and (\ref%
{1.4}), respectively, are $PT$-symmetric (note that $ix^{3}$ is not $P$%
-symmetric and $T$-symmetric, separately).

The f\/irst rigorous proof of reality and positivity of the eigenvalues of
equation (\ref{1.4}) was given in 2001 by Dorey, Dunning, and Tateo \cite{[3]} (see also \cite{[4]}).

Note that for $0<m<2$ the spectrum of (\ref{1.4}) considered in $%
L^{2}(-\infty ,\infty )$ is also discrete, however in this case only a
f\/inite number of eigenvalues are real and positive, and remaining
eigenvalues (they are of inf\/inite number) are non-real. Besides, if $m=4k$ $%
(k=1,2,\ldots )$ then for any complex value of $E$ all solutions of equation
(\ref{1.4}) belong to $L^{2}(-\infty ,\infty )$ (the so-called Weyl's
limit-circle case holds) and, therefore, all complex values of $E$ are
eigenvalues of equation~(\ref{1.4}) and hence the spectrum is not discrete
for $m=4k$ $(k=1,2,\ldots )$ if we consider the problem in $L^{2}(-\infty
,\infty ).$ In order to have in the case $m\geq 4$ a problem with discrete
spectrum Bender and Boettcher made in~\cite{[2]} an important observation
that the equation can be considered on an appropriately chosen complex
contour. Namely, it is suf\/f\/icient to consider the equation%
\begin{equation}
-\psi ^{\prime \prime }(z)-(iz)^{m}\psi (z)=E\psi (z),\qquad z\in \Gamma
\label{1.6}
\end{equation}%
together with the condition that%
\begin{equation}
\left\vert \psi (z)\right\vert \rightarrow 0 \ \text{ exponentially as }z\text{
moves of\/f to inf\/inity along }\Gamma ,  \label{1.7}
\end{equation}%
where $\Gamma $ is a contour of the form%
\[
\Gamma =\{z=x-i\left\vert x\right\vert \tan \delta :-\infty <x<\infty \}
\]%
with%
\[
\delta =\frac{\pi (m-2)}{2(m+2)}.
\]%
Note that the contour $\Gamma $ forms an angle in the lower complex $z$-plane, of value $\pi -2\delta $ with the vertex at the origin and symmetric
with respect to the imaginary axis.

Next, Mostafazadeh showed in \cite{[5]} that problem (\ref{1.6}), (\ref{1.7}) is equivalent to f\/inding solutions in $L^{2}(-\infty ,\infty )$ of the
problem%
\begin{gather}
-\psi ^{\prime \prime }(x)+\left\vert x\right\vert ^{m}\psi (x)=E\rho
(x)\psi (x),\qquad x\in (-\infty ,0)\cup (0,\infty ),  \label{1.8}
\\
\psi (0^{-})=\psi (0^{+}),\qquad \psi ^{\prime }(0^{-})=e^{2i\delta
}\psi ^{\prime }(0^{+}),  \label{1.9}
\end{gather}%
where%
\begin{equation}
\rho (x)=\left\{
\begin{array}{lll}
e^{2i\delta } & \text{if} & x<0, \\
e^{-2i\delta } & \text{if} & x>0.%
\end{array}%
\right.  \label{1.10}
\end{equation}

The main distinguishing features of problem (\ref{1.8}), (\ref{1.9}) are
that it involves a complex-valued coef\/f\/icient function $\rho (x)$ of the
form (\ref{1.10}) and that transition conditions (impulse conditions) of the
form (\ref{1.9}) are presented which also involve a complex coef\/f\/icient.
Such a problem is non-Hermitian with respect to the usual inner product (\ref%
{1.2}) of space $L^{2}(-\infty ,\infty ).$

Our aim in this paper is to construct and investigate a discrete version of
problem (\ref{1.8}), (\ref{1.9}). Discrete equations (dif\/ference equations)
form a reach f\/ield, both interesting and useful \cite{[6],[7]}. Discrete
equations arise when dif\/ferential equations are solved approximately by
discretization. On the other hand they often arise independently as
mathematical models of many practical events. Discrete equations can easily
been algorithmized to solve them on computers. There is only a small body of
work concerning discrete non-Hermitian quantum systems. Some examples are
\cite{[8],[9],[10],[11],[12],[13],[14],[15]}. Note that in \cite{[15]} the author considered a f\/inite
discrete interval version of the inf\/inite discrete interval problem (\ref{1.12}), (\ref{1.13}) formulated below, and found conditions that ensure the
reality of the eigenvalues. In the f\/inite discrete interval case with the
zero boundary conditions the problem is reduced to the eigenvalue problem
for a f\/inite-dimensional tridiagonal matrix.

Let $
\mathbb{Z}
$ denote the set of all integers. For any $l,m\in
\mathbb{Z}
$ with $l\leq m$, $[l,m]$ will denote the \textit{discrete interval }being
the set $\{l,l+1,\ldots ,m\}.$ Semi-inf\/inite intervals of the form $(-\infty
,l]$ and $[l,\infty )$ will denote the discrete sets $\{\ldots ,l-2,l-1,l\}$
and $\{l,l+1,l+2,\ldots \},$ respectively. Throughout the paper all
intervals will be discrete intervals. Let us set%
\begin{equation}
\mathbb{Z}_{0}=\mathbb{Z}
\setminus \left\{ 0,1\right\} =\left\{ \ldots ,-3,-2,-1\right\} \cup
\left\{ 2,3,4,\ldots \right\} =(-\infty ,-1]\cup \lbrack 2,\infty ).
\label{1.11}
\end{equation}%
We of\/fer a discrete version of problem (\ref{1.8}), (\ref{1.9}) to be
\begin{gather}
-\Delta ^{2}y_{n-1}+q_{n}y_{n}=\lambda \rho _{n}y_{n},\qquad n\in
\mathbb{Z}_{0},  \label{1.12}
\\
y_{-1}=y_{1},\qquad \Delta y_{-1}=e^{2i\delta }\Delta y_{1},
\label{1.13}
\end{gather}%
where $y=\left( y_{n}\right) _{n\in \mathbb{Z}
}$ is a desired solution, $\Delta $ is the forward dif\/ference operator
def\/ined by $\Delta y_{n}=y_{n+1}-y_{n}$ so that $\Delta
^{2}y_{n-1}=y_{n+1}-2y_{n}+y_{n-1},$ the coef\/f\/icients $q_{n}$ are real
numbers given for $n\in \mathbb{Z}_{0},$ $\delta \in \lbrack 0,\pi /2),$ and $\rho _{n}$ are given for $n\in
\mathbb{Z}_{0}$ by
\begin{equation}
\rho _{n}=\left\{
\begin{array}{lll}
e^{2i\delta } & \text{if} & n\leq -1, \\
e^{-2i\delta } & \text{if} & n\geq 2.%
\end{array}%
\right.  \label{1.14}
\end{equation}

One of the main results of the present paper is that if%
\begin{equation}
q_{n}\geq c>0 \qquad \text{for} \ \ n\in \mathbb{Z}_{0}  \label{1.15}
\end{equation}%
and%
\begin{equation}
\lim_{\left\vert n\right\vert \rightarrow \infty }q_{n}=\infty ,
\label{1.16}
\end{equation}%
then the spectrum of problem (\ref{1.12}), (\ref{1.13}) is discrete.

The paper is organized as follows. In Section~\ref{section2}, we choose a
suitable Hilbert space and def\/ine the main linear operators $L$, $M$, and $%
A=M^{-1}L$ related to problem (\ref{1.12}), (\ref{1.13}). Using these
operators we introduce the concept of the spectrum for problem (\ref{1.12}),
(\ref{1.13}). In Section~\ref{section3}, we demonstrate non-Hermiticity of the
operators $L$ and $A.$ In Section~\ref{section4}, we present general properties
of solutions of equations of type (\ref{1.12}), (\ref{1.13}). In Section~\ref{section5}, we construct two special solutions of problem (\ref{1.12}), (\ref{1.13}) under the condition (\ref{1.15}). Using these solutions we show in
Section~\ref{section6} that the operator $L$ is invertible and we describe the
structure of the inverse opera\-tor~$L^{-1}$. Finally, in Section~\ref{section7},
we show that the operator $L^{-1}$ is completely continuous if, in addition,
the condition (\ref{1.16}) is satisf\/ied. This fact yields the discreteness
of the spectrum of problem (\ref{1.12}), (\ref{1.13}).

\section{The concept of the spectrum for problem (\ref{1.12}), (\ref{1.13})}\label{section2}

In order to introduce the concept of the spectrum for problem (\ref{1.12}), (%
\ref{1.13}), def\/ine the Hilbert space $l_{0}^{2}$ of complex sequences $%
y=\left( y_{n}\right) _{n\in
\mathbb{Z}
_{0}}$ such that%
\[
\sum_{n\in
\mathbb{Z}
_{0}}\left\vert y_{n}\right\vert ^{2}<\infty
\]%
with the inner product and norm%
\[
\left\langle y,z\right\rangle =\sum_{n\in
\mathbb{Z}
_{0}}y_{n}\overline{z}_{n},\qquad \left\Vert y\right\Vert =\sqrt{%
\left\langle y,y\right\rangle }=\Big\{ \sum_{n\in
\mathbb{Z}
_{0}}\left\vert y_{n}\right\vert ^{2}\Big\} ^{1/2},
\]%
where $\mathbb{Z}
_{0}$ is def\/ined by (\ref{1.11}) and the bar over a complex number denotes
the complex conjugate.

Now we try to rewrite problem (\ref{1.12}), (\ref{1.13}) in the form of an
equivalent vector equation in~$l_{0}^{2}$ using appropriate operators.
Denote by $D$ the linear set of all vectors $y=\left( y_{n}\right) _{n\in
\mathbb{Z}
_{0}}\in l_{0}^{2}$ such that $\left( q_{n}y_{n}\right) _{n\in \mathbb{Z}
_{0}}\in l_{0}^{2}.$ Taking this set as the domain of $L,$ $L:D\subset
l_{0}^{2}\rightarrow l_{0}^{2}$ is def\/ined by
\[
\left( Ly\right) _{n}=-\Delta ^{2}y_{n-1}+q_{n}y_{n}\qquad \text{for} \ \ n\in
\mathbb{Z}
_{0},
\]%
where $y_{0}$ and $y_{1}$ are def\/ined from the equations
\[
y_{-1}=y_{1},\qquad \Delta y_{-1}=e^{2i\delta }\Delta y_{1}.
\]%
Note that $y_{0}$ and $y_{1}$ are needed when we evaluate $\left( Ly\right)
_{n}$ for $n=-1$ and $n=2,$ respectively:%
\begin{gather*}
(Ly)_{-1}  = -\Delta ^{2}y_{-2}+q_{-1}y_{-1}  =(y_{0}-2y_{-1}+y_{-2})+q_{-1}y_{-1},
\\
(Ly)_{2}  = -\Delta ^{2}y_{1}+q_{2}y_{2}  = (y_{3}-2y_{2}+y_{1})+q_{2}y_{2}
\end{gather*}
and (\ref{1.13}) gives for $y_{1}$, $y_{0}$ the expressions%
\begin{gather*}
y_{1}  = y_{-1}, \\
y_{0}  = y_{-1}+e^{2i\delta }(y_{2}-y_{1})  =\big(1-e^{2i\delta }\big)y_{-1}+e^{2i\delta }y_{2}.
\end{gather*}

Next, def\/ine the operator $M:l_{0}^{2}\rightarrow l_{0}^{2}$ by%
\[
\left( My\right) _{n}=\rho _{n}y_{n}\qquad \text{for} \ \ n\in
\mathbb{Z}_{0},
\]
where $\rho _{n}$ is given by (\ref{1.14}). Obviously, the adjoint $M^{\ast
} $ of $M$ is def\/ined by%
\[
(M^{\ast }y)_{n}=\overline{\rho }_{n}y_{n},\qquad n\in
\mathbb{Z}_{0},
\]%
and since $\left\vert \rho _{n}\right\vert =1,$ we get that $M$ is a unitary
operator:%
\[
MM^{\ast }=M^{\ast }M=I,
\]
where the asterisk denotes the adjoint operator and $I$ is the identity
operator.

Therefore problem (\ref{1.12}), (\ref{1.13}) can be written as
\[
Ly=\lambda My,\qquad y\in D,
\qquad \mbox{or} \qquad
M^{-1}Ly=\lambda y,\qquad y\in D.
\]%
This motivates to introduce the following def\/inition.

\begin{definition}
\label{Def2.1}By the spectrum of problem (\ref{1.12}), (\ref{1.13}) is meant
the spectrum of the operator $A=M^{-1}L$ with the domain $D$ in the space $%
l_{0}^{2}.$
\end{definition}

Remember that (see \cite{[16]}) if $A$ is a linear operator with a domain
dense in a Hilbert space, then a complex number $\lambda $ is called a
\textit{regular point }of the operator $A$ if the inverse $\left( A-\lambda
I\right) ^{-1}$ exists and represents a bounded operator def\/ined on the
whole space. All other points of the complex plane comprise the \textit{spectrum} of the operator $A.$ Obviously the eigenvalues $\lambda $ of an
operator belong to its spectrum, since the operator $\left( A-\lambda
I\right) ^{-1}$ does not exist for such points (the operator $A-\lambda I$
is not one-to-one). The set of all eigenvalues is called the \textit{point
spectrum} of the operator. The spectrum of the operator $A$ is said to be
\textit{discrete }if it consists of a denumerable (i.e., at most countable)
set of eigenvalues with no f\/inite point of accumulation.

A linear operator acting in a Hilbert space and def\/ined on the whole space
is called \textit{completely continuous }if it maps bounded sets into
relatively compact sets (a set is called \textit{relatively compact} if
every inf\/inite subset of this set has a limit point in the space, that may
not belong to the set). Any completely continuous operator is bounded and
hence its spectrum is a compact subset of the complex plane. As is well
known~\cite{[16]}, every nonzero point of the spectrum of a completely
continuous operator is an eigenvalue of f\/inite multiplicity (that is, to
each eigenvalue there correspond only a f\/inite number of linearly
independent eigenvectors); the set of eigenvalues is at most countable and
can have only one accumulation point $\lambda =0.$ It follows that if a
linear operator $A$ with a domain dense in a Hilbert space is invertible and
its inverse $A^{-1}$ is completely continuous, then the spectrum of $A$ is
discrete.

In this paper we show that the operator $L$
 is invertible under the
condition (\ref{1.15}) and that its inverse $L^{-1}$ is a completely
continuous operator if, in addition, the condition (\ref{1.16}) is
satisf\/ied. This implies that the operator $A=M^{-1}L$ is invertible and $%
A^{-1}=L^{-1}M$ is a completely continuous operator. Hence the spectrum of
the operator $A$ is discrete.

\section[Non-Hermiticity of the operators  $L$ and  $A$]{Non-Hermiticity of the operators  $\boldsymbol{L}$ and  $\boldsymbol{A}$}\label{section3}

Let $(f_{k})$ be a given complex sequence, where $k\in
\mathbb{Z}.$ The forward and backward dif\/ference operators $\Delta $ and $\nabla $ are
def\/ined by%
\[
\Delta f_{k}=f_{k+1}-f_{k}\qquad \text{and} \qquad \nabla f_{k}=f_{k}-f_{k-1},
\]%
respectively. We easily see that%
\begin{gather*}
\nabla f_{k}=\Delta f_{k-1},
\\
\Delta ^{2}f_{k}=\Delta (\Delta f_{k})=f_{k+2}-2f_{k+1}+f_{k},
\\
\nabla ^{2}f_{k}=\nabla (\nabla f_{k})=f_{k}-2f_{k-1}+f_{k-2},
\\
\Delta \nabla f_{k}=f_{k+1}-2f_{k}+f_{k-1}=\nabla \Delta f_{k}=\Delta
^{2}f_{k-1}=\nabla ^{2}f_{k+1}.
\end{gather*}
For any integers $a,b\in
\mathbb{Z}
$ with $a<b$ we have the summation by parts formulas%
\begin{gather}
\sum_{k=a}^{b}(\Delta f_{k})g_{k}=f_{k+1}g_{k}\big|_{a-1}^{b}-\sum_{k=a}^{b}f_{k}(\nabla
g_{k})=f_{b+1}g_{b}-f_{a}g_{a-1}-\sum_{k=a}^{b}f_{k}(\nabla g_{k}),
\nonumber\\
\sum_{k=a}^{b}(\nabla f_{k})g_{k}=f_{k}g_{k+1}\big|_{a-1}^{b}-\sum_{k=a}^{b}f_{k}(\Delta
g_{k})=f_{b}g_{b+1}-f_{a-1}g_{a}-\sum_{k=a}^{b}f_{k}(\Delta g_{k}),
\label{3.1}
\\
\sum_{k=a}^{b}(\Delta \nabla f_{k})g_{k}=(\Delta f_{k})g_{k}\big|_{a-1}^{b}-\sum_{k=a}^{b}(\nabla f_{k})(\nabla g_{k}),  \label{3.2}
\\
\sum_{k=a}^{b}(\Delta \nabla f_{k})g_{k}=(\Delta f_{k})g_{k+1}\big|_{a-1}^{b}-\sum_{k=a}^{b}(\Delta f_{k})(\Delta g_{k}),  \nonumber %\label{3.3}
\\
\sum_{k=a}^{b}[(\Delta \nabla f_{k})g_{k}-f_{k}(\Delta \nabla
g_{k})]=[(\Delta f_{k})g_{k}-f_{k}(\Delta g_{k})]_{a-1}^{b}
\nonumber\\
\qquad{}
=[(\Delta f_{b})g_{b}-f_{b}(\Delta g_{b})]-[(\Delta
f_{a-1})g_{a-1}-f_{a-1}(\Delta g_{a-1})].  \label{3.4}
\end{gather}

\begin{theorem}
\label{Th3.1}Let $\delta \in \lbrack 0,\pi /2).$ If $\delta =0,$ then the
operator $L$ is Hermitian:%
\[
\left\langle Ly,z\right\rangle =\left\langle y,Lz\right\rangle \qquad \text{for all} \ \ y,z\in D.
\]%
But if $\delta \neq 0,$ then the operator $L$ is not Hermitian.
\end{theorem}

\begin{proof}
Using formula (\ref{3.4}) and equation%
\begin{equation*}
\left( Ly\right) _{n}=-\Delta ^{2}y_{n-1}+q_{n}y_{n}=-\Delta \nabla
y_{n}+q_{n}y_{n}\qquad \text{for} \ \ n\in
\mathbb{Z}_{0},  %\label{3.5}
\end{equation*}%
where $y_{0}$ and $y_{1}$ are def\/ined from the equations%
\begin{equation}
y_{-1}=y_{1},\qquad \Delta y_{-1}=e^{2i\delta }\Delta y_{1},
\label{3.6}
\end{equation}
and taking into account that for any $y=\left( y_{n}\right) _{n\in \mathbb{Z}
_{0}}\in l_{0}^{2}$ we have $y_{n}\rightarrow 0,$ $\Delta y_{n}\rightarrow 0$
as $\left\vert n\right\vert \rightarrow \infty ,$ we get for all $y,z\in D$,
\begin{gather*}
\left\langle Ly,z\right\rangle -\left\langle y,Lz\right\rangle =-\sum_{n\in
\mathbb{Z}
_{0}}\left[ (\Delta \nabla y_{n})\overline{z}_{n}-y_{n}(\Delta \nabla
\overline{z}_{n})\right]
\\
\phantom{\left\langle Ly,z\right\rangle -\left\langle y,Lz\right\rangle}{}
=-\sum_{-\infty }^{n=-1}\left[ (\Delta \nabla y_{n})\overline{z}%
_{n}-y_{n}(\Delta \nabla \overline{z}_{n})\right] -\sum_{n=2}^{\infty }\left[
(\Delta \nabla y_{n})\overline{z}_{n}-y_{n}(\Delta \nabla \overline{z}_{n})%
\right]
\\
\phantom{\left\langle Ly,z\right\rangle -\left\langle y,Lz\right\rangle}{}
=-[(\Delta y_{-1})\overline{z}_{-1}-y_{-1}(\Delta \overline{z}%
_{-1})]+[(\Delta y_{1})\overline{z}_{1}-y_{1}(\Delta \overline{z}_{1})]
\\
\phantom{\left\langle Ly,z\right\rangle -\left\langle y,Lz\right\rangle}{}
=-[(\Delta y_{-1})\overline{z}_{-1}-y_{-1}(\Delta \overline{z}%
_{-1})]+\big[e^{-2i\delta }(\Delta y_{-1})\overline{z}_{-1}-y_{-1}e^{2i\delta
}(\Delta \overline{z}_{-1})\big]
\\
\phantom{\left\langle Ly,z\right\rangle -\left\langle y,Lz\right\rangle}{}
=\big(e^{-2i\delta }-1\big)(\Delta y_{-1})\overline{z}_{-1}-\big(e^{2i\delta
}-1\big)y_{-1}(\Delta \overline{z}_{-1}).
\end{gather*}
Thus,%
\begin{equation}
\left\langle Ly,z\right\rangle -\left\langle y,Lz\right\rangle
=\big(e^{-2i\delta }-1\big)(\Delta y_{-1})\overline{z}_{-1}-\big(e^{2i\delta
}-1\big)y_{-1}(\Delta \overline{z}_{-1}),  \label{3.7}
\end{equation}%
for all $y,z\in D.$ Formula (\ref{3.7}) shows that if $\delta =0,$ then the
operator $L$ is Hermitian:%
\[
\left\langle Ly,z\right\rangle =\left\langle y,Lz\right\rangle \qquad \text{for all} \ \ y,z\in D.
\]%
The same formula shows that if $\delta \neq 0$ (recall that $\delta \in
\lbrack 0,\pi /2)$), then the operator $L$ is not Hermitian:%
\[
\left\langle Ly,z\right\rangle \neq \left\langle y,Lz\right\rangle \qquad \text{for some} \ \ y,z\in D.
\]%
The theorem is proved.
\end{proof}

\begin{theorem}
\label{Th3.2}Let $\delta \in \lbrack 0,\pi /2).$ If $\delta =0,$ then the
operator $A=M^{-1}L$ is Hermitian:%
\[
\left\langle Ay,z\right\rangle =\left\langle y,Az\right\rangle \qquad \text{for all} \ \ y,z\in D.
\]%
But if $\delta \neq 0,$ then the operator $A$ is not Hermitian.
\end{theorem}

\begin{proof}
We have for any $y,z\in D,$%
\begin{gather*}
\left\langle Ay,z\right\rangle -\left\langle y,Az\right\rangle =\left\langle
M^{-1}Ly,z\right\rangle -\left\langle y,M^{-1}Lz\right\rangle =\left\langle
Ly,Mz\right\rangle -\left\langle My,Lz\right\rangle
\\
\phantom{\left\langle Ay,z\right\rangle -\left\langle y,Az\right\rangle}{}
=\sum_{n\in
\mathbb{Z}
_{0}}[-(\Delta \nabla y_{n})+q_{n}y_{n}]\overline{\rho _{n}z_{n}}-\sum_{n\in
\mathbb{Z}
_{0}}\rho _{n}y_{n}[-(\Delta \nabla \overline{z}_{n})+q_{n}\overline{z}_{n}]
\\
\phantom{\left\langle Ay,z\right\rangle -\left\langle y,Az\right\rangle}{}
=-\sum_{n\in
\mathbb{Z}
_{0}}[(\Delta \nabla y_{n})\overline{\rho _{n}z_{n}}-\rho _{n}y_{n}(\Delta
\nabla \overline{z}_{n})]+\sum_{n\in
\mathbb{Z}
_{0}}(\overline{\rho }_{n}-\rho _{n})q_{n}y_{n}\overline{z}_{n}.
\end{gather*}
Next, from%
\[
\rho _{n}=\left\{
\begin{array}{lll}
e^{2i\delta } & \text{if} & n\leq -1, \\
e^{-2i\delta } & \text{if} & n\geq 2,
\end{array}%
\right. \qquad \text{and}\qquad \overline{\rho }_{n}=\left\{
\begin{array}{lll}
e^{-2i\delta } & \text{if} & n\leq -1, \\
e^{2i\delta } & \text{if} & n\geq 2,
\end{array}%
\right.
\]
we f\/ind%
\[
\overline{\rho }_{n}-\rho _{n}=\left\{
\begin{array}{lll}
-2i\sin 2\delta & \text{if} & n\leq -1, \\
2i\sin 2\delta & \text{if} & n\geq 2,
\end{array}%
\right.
\]%
so that%
\[
\sum_{n\in
\mathbb{Z}
_{0}}(\overline{\rho }_{n}-\rho _{n})q_{n}y_{n}\overline{z}_{n}=-2i\sin
2\delta \sum_{-\infty }^{n=-1}q_{n}y_{n}\overline{z}_{n}+2i\sin 2\delta
\sum_{n=2}^{\infty }q_{n}y_{n}\overline{z}_{n}.
\]
Besides, using formula (\ref{3.2}) and equations in (\ref{3.6}), we obtain%
\begin{gather*}
-\sum_{n\in
\mathbb{Z}
_{0}}[(\Delta \nabla y_{n})\overline{\rho _{n}z_{n}}-\rho _{n}y_{n}(\Delta
\nabla \overline{z}_{n})]
=-e^{-2i\delta }\sum_{-\infty }^{n=-1}(\Delta \nabla y_{n})\overline{z}%
_{n}-e^{2i\delta }\sum_{n=2}^{\infty }(\Delta \nabla y_{n})\overline{z}%
_{n}\\
\quad \qquad{}+e^{2i\delta }\sum_{-\infty }^{n=-1}y_{n}(\Delta \nabla \overline{z}%
_{n})+e^{-2i\delta }\sum_{n=2}^{\infty }y_{n}(\Delta \nabla \overline{z}_{n})
\\
\qquad{}=-e^{-2i\delta }(\Delta y_{-1})\overline{z}_{-1}+e^{-2i\delta }\sum_{-\infty
}^{n=-1}(\nabla y_{n})(\nabla \overline{z}_{n})+e^{2i\delta }(\Delta y_{1})%
\overline{z}_{1}+e^{2i\delta }\sum_{n=2}^{\infty }(\nabla y_{n})(\nabla
\overline{z}_{n})
\\
\qquad\quad{}+e^{2i\delta }y_{-1}(\Delta \overline{z}_{-1})-e^{2i\delta }\sum_{-\infty
}^{n=-1}(\nabla y_{n})(\nabla \overline{z}_{n})-e^{-2i\delta }y_{1}(\Delta
\overline{z}_{1})-e^{-2i\delta }\sum_{n=2}^{\infty }(\nabla y_{n})(\nabla
\overline{z}_{n})
\\
\qquad{}=\big(1-e^{-2i\delta }\big)(\Delta y_{-1})\overline{z}_{-1}-\big(1-e^{2i\delta
}\big)y_{-1}(\Delta \overline{z}_{-1})
\\
\quad\qquad{}-2i\sin 2\delta \sum_{-\infty }^{n=-1}(\nabla y_{n})(\nabla \overline{z}%
_{n})+2i\sin 2\delta \sum_{n=2}^{\infty }(\nabla y_{n})(\nabla \overline{z}%
_{n}).
\end{gather*}
Thus,%
\begin{gather}
\left\langle Ay,z\right\rangle -\left\langle y,Az\right\rangle
=\big(1-e^{-2i\delta }\big)(\Delta y_{-1})\overline{z}_{-1}-\big(1-e^{2i\delta
}\big)y_{-1}(\Delta \overline{z}_{-1})\nonumber
\\
\qquad{}
-2i\sin 2\delta \sum_{-\infty }^{n=-1}[(\nabla y_{n})(\nabla \overline{z}%
_{n})+q_{n}y_{n}\overline{z}_{n}]+2i\sin 2\delta \sum_{n=2}^{\infty
}[(\nabla y_{n})(\nabla \overline{z}_{n})+q_{n}y_{n}\overline{z}_{n}].\label{3.8}
\end{gather}%
Formula (\ref{3.8}) shows that if $\delta =0,$ then the operator $A$ is
Hermitian and if $\delta \neq 0,$ then $A$ is not Hermitian.
\end{proof}

\begin{remark}
In the case $\delta =0$ we have $A=L$.
\end{remark}

\section[Second order linear difference equations with impulse]{Second order linear dif\/ference equations with impulse}\label{section4}

Consider the second order linear homogeneous dif\/ference equation with impulse%
\begin{gather}
-\Delta ^{2}y_{n-1}+p_{n}y_{n}=0,\qquad n\in \mathbb{Z}_{0}=%
\mathbb{Z}
\backslash \left\{ 0,1\right\} =(-\infty ,-1]\cup \lbrack 2,\infty ),
\label{4.1}
\\
y_{-1}=d_{1}y_{1},\qquad \Delta y_{-1}=d_{2}\Delta y_{1},  \label{4.2}
\end{gather}%
where $y=\left( y_{n}\right) $ with $n\in \mathbb{Z}
$ is a desired solution, the coef\/f\/icients $p_{n}$ are complex numbers given
for $n\in
\mathbb{Z}
_{0}$; $d_{1}$, $d_{2}$ presented in the ``impulse conditions'' (transition
conditions) in (\ref{4.2}) are nonzero complex numbers.

Using the def\/inition of $\Delta $-derivative we can rewrite problem (\ref{4.1}), (\ref{4.2}) in the form%
\begin{gather}
-y_{n-1}+\widetilde{p}_{n}y_{n}-y_{n+1}=0,\qquad n\in (-\infty ,-1]\cup
\lbrack 2,\infty ),  \label{4.3}
\\
y_{-1}=d_{1}y_{1},\qquad y_{0}-y_{-1}=d_{2}(y_{2}-y_{1}),  \label{4.4}
\end{gather}%
where%
\[
\widetilde{p}_{n}=p_{n}+2,\qquad n\in (-\infty ,-1]\cup \lbrack
2,\infty ).
\]

\begin{theorem}
\label{Th4.1}Let $n_{0}$ be a fixed point in $\mathbb{Z}$ and $c_{0}$, $c_{1}$ be given complex numbers. Then problem \eqref{4.1}, \eqref{4.2} has a unique solution $(y_{n})$, $n\in
\mathbb{Z}$, such that%
\begin{equation}
y_{n_{0}}=c_{0},\qquad \Delta y_{n_{0}}=c_{1},\qquad \text{that is,}\qquad
y_{n_{0}}=c_{0},\qquad y_{n_{0}+1}=c_{0}+c_{1}=c_{1}^{\prime }.
\label{4.5}
\end{equation}
\end{theorem}

\begin{proof}
First assume that $n_{0}\in (-\infty ,-1].$ We can rewrite equation (\ref{4.3}) in the form%
\begin{equation}
y_{n-1}=\widetilde{p}_{n}y_{n}-y_{n+1}=0,\qquad n\in (-\infty ,-1]\cup
\lbrack 2,\infty )  \label{4.6}
\end{equation}%
as well as in the form%
\begin{equation}
y_{n+1}=\widetilde{p}_{n}y_{n}-y_{n-1}=0,\qquad n\in (-\infty ,-1]\cup
\lbrack 2,\infty ).  \label{4.7}
\end{equation}%
Using the initial conditions (\ref{4.5}) we f\/ind, recurrently (step by
step), $y_{n}$ for $n\leq n_{0}+1$ uniquely from (\ref{4.6}) and for $%
n_{0}+2\leq n\leq -1$ uniquely from (\ref{4.7}). Then we f\/ind $y_{1}$ and $%
y_{2}$ from the transition conditions (\ref{4.4}) and then we f\/ind $y_{n}$
for $n\geq 3$ uniquely from (\ref{4.7}).

In the case $n_{0}\in \lbrack 1,\infty )$ we are reasoning similarly; using
equations (\ref{4.6}), (\ref{4.7}) we f\/irst f\/ind~$y_{n}$ uniquely for $n\geq
1$ and then using the transition conditions (\ref{4.4}) we pass to the
interval $(-\infty ,-1].$

Finally, if $n_{0}=0,$ then we f\/ind $y_{0}$ and $y_{1}$ uniquely from the
initial conditions (\ref{4.5}) with $n_{0}=0$. Then we f\/ind $y_{-1}$ and $%
y_{2}$ from the transition conditions (\ref{4.4}). Next, solving equation~(\ref{4.6}) at f\/irst on $(-\infty ,-1]$ we f\/ind $y_{n}$ uniquely for $n\in
(-\infty ,-2]$ and then solving (\ref{4.7}) on $[2,\infty )$ we f\/ind $y_{n}$
uniquely for $n\in \lbrack 3,\infty ).$
\end{proof}

\begin{definition}
\label{Def4.2}For two sequences $y=(y_{n})$ and $z=(z_{n})$ with $n\in
\mathbb{Z}
$, we def\/ine their Wronskian~by%
\begin{gather*}
W_{n}(y,z) = y_{n}\Delta z_{n}-(\Delta y_{n})z_{n}  = y_{n}z_{n+1}-y_{n+1}z_{n},\qquad n\in
\mathbb{Z}
.
\end{gather*}
\end{definition}

\begin{theorem}
\label{Th4.3} The Wronskian of any two solutions $y$ and $z$ of problem \eqref{4.1}, \eqref{4.2} is constant on each of the intervals $(-\infty ,-1]$ and
$[1,\infty )$:
\begin{equation}
W_{n}(y,z)=\left\{
\begin{array}{lll}
\omega ^{-} & \text{if} & n\in (-\infty ,-1], \\
\omega ^{+} & \text{if} & n\in \lbrack 1,\infty ).%
\end{array}%
\right.  \label{4.8}
\end{equation}%
In addition,
\begin{equation}
\omega ^{-}=d_{1}d_{2}\omega ^{+}  \label{4.9}
\end{equation}%
and%
\begin{equation}
W_{0}(y,z)=-d_{2}\omega ^{+}.  \label{4.10}
\end{equation}
\end{theorem}

\begin{proof}
Suppose that $y=(y_{n})$ and $z=(z_{n}),$ where $n\in
\mathbb{Z}
,$ are solutions of (\ref{4.1}), (\ref{4.2}). Let us compute the $\Delta $%
-derivative of $W_{n}(y,z)$. Using the product rule for $\Delta $-derivative%
\begin{gather*}
\Delta (f_{n}g_{n}) = (\Delta f_{n})g_{n}+f_{n+1}\Delta g_{n}
= f_{n}\Delta g_{n}+(\Delta f_{n})g_{n+1},
\end{gather*}
we have
\begin{gather*}
\Delta W_{n}(y,z)=\Delta \left[ y_{n}\Delta z_{n}-(\Delta y_{n})z_{n}\right]
=(\Delta y_{n})\Delta z_{n}+y_{n+1}\Delta ^{2}z_{n}-(\Delta y_{n})\Delta
z_{n}-(\Delta ^{2}y_{n})z_{n+1}
\\
\phantom{\Delta W_{n}(y,z)}{}
=y_{n+1}\Delta ^{2}z_{n}-(\Delta ^{2}y_{n})z_{n+1}.
\end{gather*}
Further, since $y_{n}$ and $z_{n}$ are solutions of (\ref{4.1}), (\ref{4.2}),%
\begin{gather*}
\Delta ^{2}y_{n}=p_{n+1}y_{n+1},\qquad n\in (-\infty ,-2]\cup \lbrack
1,\infty ),
\\
\Delta ^{2}z_{n}=p_{n+1}z_{n+1},\qquad n\in (-\infty ,-2]\cup \lbrack
1,\infty ).
\end{gather*}
Therefore%
\[
\Delta W_{n}(y,z)=0\qquad \text{for} \ \ n\in (-\infty ,-2]\cup \lbrack
1,\infty ).
\]%
The latter implies that $W_{n}(y,z)$ is constant on $(-\infty ,-1]$ and on $%
[1,\infty ).$ Thus we have (\ref{4.8}), where $\omega ^{-}$ and $\omega ^{+}$
are some constants (depending on the solutions $y$ and $z$).

Next using (\ref{4.8}) and the impulse conditions in (\ref{4.2}) for $y_{n}$
and $z_{n},$ we have{\samepage
\begin{gather*}
\omega ^{-}=W_{-1}(y,z)=y_{-1}\Delta z_{-1}-(\Delta y_{-1})z_{-1}
=d_{1}d_{2}[y_{1}\Delta z_{1}-(\Delta
y_{1})z_{1}]\\
\phantom{\omega ^{-}}{} =d_{1}d_{2}W_{1}(y,z)=d_{1}d_{2}\omega ^{+},
\end{gather*}
so that (\ref{4.9}) is established.}

Finally, from the impulse conditions in (\ref{4.2}) we f\/ind that%
\[
y_{0}=(d_{1}-d_{2})y_{1}+d_{2}y_{2}.
\]%
Substituting this expression for $y_{0}$ and $z_{0}$ into
\[
W_{0}(y,z)=y_{0}z_{1}-y_{1}z_{0},
\]%
we get%
\[
W_{0}(y,z)=-d_{2}W_{1}(y,z)=-d_{2}\omega ^{+}.
\]%
Therefore (\ref{4.10}) is also proved.
\end{proof}

\begin{corollary}
\label{Cor4.4}If $y$ and $z$ are two solutions of \eqref{4.1}, \eqref{4.2},
then either $W_{n}(y,z)=0$ for all $n\in
\mathbb{Z}$ or $W_{n}(y,z)\neq 0$ for all $n\in
\mathbb{Z}.$
\end{corollary}

By using Theorem \ref{Th4.1}, the following two theorems can be proved in
exactly the same way when equation (\ref{4.1}) does not include any impulse
conditions \cite{[6]}.

\begin{theorem}
\label{Th4.5}Any two solutions of \eqref{4.1}, \eqref{4.2} are linearly
independent if and only if their Wronskian is not zero.
\end{theorem}

\begin{theorem}
\label{Th4.6}Problem \eqref{4.1}, \eqref{4.2} has two linearly independent
solutions and every solution of~\eqref{4.1},~\eqref{4.2} is a linear
combination of these solutions.
\end{theorem}

We say that $y=(y_{n})$ and $z=(z_{n}),$ where $n\in
%TCIMACRO{\U{2124} }%
%BeginExpansion
\mathbb{Z}
%EndExpansion
,$ form a \textit{fundamental set }(or \textit{fundamental system}) of
solutions for (\ref{4.1}), (\ref{4.2}) provided that they are solutions of (%
\ref{4.1}), (\ref{4.2}) and their Wronskian is not zero.

Let us consider the nonhomogeneous equation%
\begin{equation}
-\Delta ^{2}y_{n-1}+p_{n}y_{n}=h_{n},\qquad n\in (-\infty ,-1]\cup
\lbrack 2,\infty ),  \label{4.11}
\end{equation}%
with the impulse conditions%
\begin{equation}
y_{-1}=d_{1}y_{1},\qquad \Delta y_{-1}=d_{2}\Delta y_{1},  \label{4.12}
\end{equation}%
where $h_{n}$ is a complex sequence def\/ined for $n\in (-\infty ,-1]\cup
\lbrack 2,\infty ).$ We will extend $h_{n}$ to the values $n=0$ and $n=1$ by
setting%
\begin{equation}
h_{0}=h_{1}=0.  \label{4.13}
\end{equation}

\begin{theorem}
\label{Th4.7}Suppose that $u=(u_{n})$ and $v=(v_{n})$ form a fundamental set
of solutions of the homogeneous problem \eqref{4.1}, \eqref{4.2}. Then a
general solution of the corresponding nonhomogeneous problem \eqref{4.11}, \eqref{4.12} is given by%
\[
y_{n}=c_{1}u_{n}+c_{2}v_{n}+x_{n},\qquad n\in
\mathbb{Z},
\]%
where $c_{1}$, $c_{2}$ are arbitrary constants and%
\begin{equation}
x_{n}=\left\{
\begin{array}{lll}
\displaystyle -\sum_{s=n}^{0}\frac{u_{n}v_{s}-u_{s}v_{n}}{W_{s}(u,v)}h_{s} & \text{if} &
n\leq 0, \vspace{1mm} \\
\displaystyle \sum_{s=1}^{n}\frac{u_{n}v_{s}-u_{s}v_{n}}{W_{s}(u,v)}h_{s} & \text{if} &
n\geq 1.%
\end{array}%
\right.  \label{4.14}
\end{equation}
\end{theorem}

\begin{proof}
Taking into account (\ref{4.13}) it is not dif\/f\/icult to verify that the
sequence $x_{n}$ def\/ined by~(\ref{4.14}) is a particular solution of (\ref%
{4.11}), (\ref{4.12}), namely, $x_{n}$ satisf\/ies equation (\ref{4.11}) and
the conditions%
\[
x_{-1}=\Delta x_{-1}=0,\qquad x_{1}=\Delta x_{1}=0.
\]%
This implies that the statement of the theorem is true.
\end{proof}

\section{Two special solutions}\label{section5}

Consider the homogeneous problem%
\begin{gather}
-\Delta ^{2}y_{n-1}+q_{n}y_{n}=0,\qquad n\in
\mathbb{Z}
_{0}=
\mathbb{Z}
\backslash \{0,1\},  \label{5.1}
\\
y_{-1}=y_{1},\qquad \Delta y_{-1}=e^{2i\delta }\Delta y_{1},
\label{5.2}
\end{gather}%
where $\delta \in \lbrack 0,\pi /2)$ is a f\/ixed real number and%
\begin{equation}
q_{n}\geq c>0\qquad \text{for} \qquad n\in
\mathbb{Z}
_{0}.  \label{5.3}
\end{equation}

In this section we show that under the condition (\ref{5.3}) problem (\ref%
{5.1}), (\ref{5.2}) has two linearly independent solutions $\psi =(\psi
_{n}) $ and $\chi =(\chi _{n}),$ where $n\in
\mathbb{Z}
,$ such that%
\begin{equation}
\sum_{n=0}^{\infty }\left\vert \psi _{n}\right\vert ^{2}<\infty \qquad \text{and} \qquad \sum_{-\infty }^{n=0}\left\vert \chi _{n}\right\vert ^{2}<\infty .
\label{5.4}
\end{equation}%
These solutions will allow us to f\/ind the inverse $L^{-1}$ of the operator $%
L $ introduced above in Section~\ref{section2} and investigate the properties of
$L^{-1}.$

First we derive two simple useful formulas related to the nonhomogeneous
problem%
\begin{gather}
-\Delta ^{2}y_{n-1}+q_{n}y_{n}=f_{n},\qquad n\in
\mathbb{Z}
_{0},  \label{5.5}
\\
y_{-1}=y_{1},\qquad \Delta y_{-1}=e^{2i\delta }\Delta y_{1},
\label{5.6}
\end{gather}
where $(q_{n})$ is a real sequence with $n\in
\mathbb{Z}
_{0},$ and $\delta \in \lbrack 0,\pi /2);$ $(f_{n})$ is a complex sequence
with $n\in
\mathbb{Z}
_{0}.$

\begin{lemma}
\label{Lem5.1}Let $y=(y_{n})$ with $n\in
\mathbb{Z}
$ be a solution of problem \eqref{5.5}, \eqref{5.6} and $a,$ $b$ be any
integers such that $a\leq -1$ and $b\geq 2$. Then the following formulas
hold:
\begin{gather}
\sum_{n=2}^{b}\big( \left\vert \Delta y_{n}\right\vert ^{2}+q_{n}\left\vert
y_{n}\right\vert ^{2}\big) =(\Delta y_{n})\overline{y}_{n+1}\big|
_{1}^{b}+\sum_{n=2}^{b}f_{n}\overline{y}_{n},  \label{5.7}
\\
\sum_{n=a}^{-1}\big( \left\vert \Delta y_{n}\right\vert
^{2}+q_{n}\left\vert y_{n}\right\vert ^{2}\big) =(\Delta y_{n})\overline{y}%
_{n+1}\big|_{a-1}^{-1}+\sum_{n=a}^{-1}f_{n}\overline{y}_{n}.  \label{5.8}
\end{gather}
\end{lemma}

\begin{proof}
To prove (\ref{5.7}), multiply equation (\ref{5.5}) by $\overline{y}_{n}$
and sum from $n=2$ to $n=b$:
\[
-\sum_{n=2}^{b}(\Delta ^{2}y_{n-1})\overline{y}_{n}+\sum_{n=2}^{b}q_{n}\left%
\vert y_{n}\right\vert ^{2}=\sum_{n=2}^{b}f_{n}\overline{y}_{n}.
\]%
Next, applying the summation by parts formula (\ref{3.1}) we get that%
\[
-\sum_{n=2}^{b}(\Delta ^{2}y_{n-1})\overline{y}_{n}=-\sum_{n=2}^{b}(\nabla
\Delta y_{n})\overline{y}_{n}=-(\Delta y_{n})\overline{y}_{n+1}\big|
_{1}^{b}+\sum_{n=2}^{b}\left\vert \Delta y_{n}\right\vert ^{2}.
\]%
Therefore the formula (\ref{5.7}) follows.

The formula (\ref{5.8}) can be proved in a similar way.
\end{proof}

\begin{theorem}
\label{Th5.2}Under the condition \eqref{5.3} problem \eqref{5.1}, \eqref{5.2} has two linearly independent solutions $\psi =(\psi _{n})$ and $\chi =(\chi
_{n})$ with $n\in
\mathbb{Z}
$, possessing the properties stated in \eqref{5.4}.
\end{theorem}

\begin{proof}
Denote by $\varphi =(\varphi _{n})$ and $\theta =(\theta _{n}),$ where $n\in
\mathbb{Z}
,$ solutions of problem (\ref{5.1}), (\ref{5.2}) satisfying the initial
conditions%
\begin{gather}
\varphi _{1}=1,\qquad \Delta \varphi _{1}=-1,  \label{5.9}
\\
\theta _{1}=1,\qquad \Delta \theta _{1}=0.  \label{5.10}
\end{gather}%
Such solutions exist and are unique by Theorem~\ref{Th4.1}. It follows from (%
\ref{5.9}), (\ref{5.10}) that $\varphi _{2}=0,$ $\theta _{2}=1.$ According
to Theorem \ref{Th4.3} we f\/ind that%
\begin{equation}
W_{0}(\varphi ,\theta )=-e^{2i\delta },\qquad W_{n}(\varphi ,\theta
)=\left\{
\begin{array}{lll}
e^{2i\delta } & \text{if} & n\leq -1, \\
1 & \text{if} & n\geq 1.%
\end{array}%
\right.  \label{5.11}
\end{equation}%
Therefore $W_{n}(\varphi ,\theta )\neq 0$ and by Theorem \ref{Th4.5} the
solutions $\varphi $ and $\theta $ are linearly independent.

We seek the desired solution $\psi =(\psi _{n})$ of problem (\ref{5.1}), (\ref{5.2}) in the form%
\begin{equation}
\psi _{n}=\varphi _{n}+v\theta _{n},\qquad n\in
\mathbb{Z}
,  \label{5.12}
\end{equation}%
where $v$ is a complex constant which we will choose.

Take an arbitrary integer $b\geq 2$. Applying (\ref{5.7}) to%
\begin{gather*}
-\Delta ^{2}\psi _{n-1}+q_{n}\psi _{n}=0,\qquad n\in
\mathbb{Z}
_{0},
\\
\psi _{-1}=\psi _{1},\qquad \Delta \psi _{-1}=e^{2i\delta }\Delta \psi
_{1},
\end{gather*}
we get%
\begin{equation*}
\sum_{n=2}^{b}\big( \left\vert \Delta \psi _{n}\right\vert
^{2}+q_{n}\left\vert \psi _{n}\right\vert ^{2}\big) =(\Delta \psi _{n})%
\overline{\psi }_{n+1}\big|_{1}^{b}.  %\label{5.13}
\end{equation*}%
Since $\Delta \psi _{1}=-1$ and $\psi _{2}=v$, by (\ref{5.12}) and (\ref{5.9}), (\ref{5.10}), hence%
\[
\sum_{n=2}^{b}\big( \left\vert \Delta \psi _{n}\right\vert
^{2}+q_{n}\left\vert \psi _{n}\right\vert ^{2}\big) =(\Delta \psi _{b})%
\overline{\psi }_{b+1}+\overline{v}.
\]%
Multiply the latter equality by $e^{i\delta }$ and take then the real part
of both sides to get%
\begin{equation}
(\cos \delta )\sum_{n=2}^{b}\big( \left\vert \Delta \psi _{n}\right\vert
^{2}+q_{n}\left\vert \psi _{n}\right\vert ^{2}\big) ={\rm Re}\big\{e^{i\delta
}(\Delta \psi _{b})\overline{\psi }_{b+1}\big\}+{\rm Re}\big(ve^{-i\delta }\big).
\label{5.14}
\end{equation}

Now we choose $v$ so that to have%
\begin{equation}
{\rm Re}\big\{e^{i\delta }(\Delta \psi _{b})\overline{\psi }_{b+1}\big\}=0.
\label{5.15}
\end{equation}%
Since%
\[
{\rm Re}\big\{e^{i\delta }(\Delta \psi _{b})\overline{\psi }_{b+1}\big\}=\left\vert
\psi _{b+1}\right\vert ^{2}{\rm Re}\left\{ e^{i\delta }\frac{\Delta \psi
_{b}}{\psi _{b+1}}\right\} ,
\]%
it is suf\/f\/icient for (\ref{5.15}) to have%
\begin{equation}
{\rm Re}\left\{ e^{i\delta }\frac{\Delta \psi _{b}}{\psi _{b+1}}\right\} =0.
\label{5.16}
\end{equation}%
Note that $\psi _{b}$ cannot be zero for any two successive values of $b$
(otherwise $\psi _{n}$ would be identically zero by the uniqueness theorem
for solution, that is not true since $\varphi $ and $\theta $ are linearly
independent). Therefore $\psi _{b}\neq 0$ for inf\/initely many values of $b.$

Under the condition (\ref{5.16}) the equation (\ref{5.14}) becomes%
\begin{equation*}
(\cos \delta )\sum_{n=2}^{b}\big( \left\vert \Delta \psi _{n}\right\vert
^{2}+q_{n}\left\vert \psi _{n}\right\vert ^{2}\big) ={\rm Re}
\big(ve^{-i\delta }\big).  %\label{5.17}
\end{equation*}%
The condition (\ref{5.16}) can be written as%
\begin{equation}
e^{i\delta }\frac{\Delta \varphi _{b}+v\Delta \theta _{b}}{\varphi
_{b+1}+v\theta _{b+1}}=\beta ,  \label{5.18}
\end{equation}%
where $\beta $ is a pure imaginary number ($\beta =it,$ $t\in
\mathbb{R}
)$. Note that
\begin{equation}
\varphi _{b+1}\Delta \theta _{b}-(\Delta \varphi _{b})\theta _{b+1}=-\varphi
_{b+1}\theta _{b}+\varphi _{b}\theta _{b+1}=W_{b}(\varphi ,\theta )=1\neq 0
\label{5.19}
\end{equation}
by (\ref{5.11}). Therefore (\ref{5.18}) def\/ines a linear-fractional
transformation of the complex $v$-plane onto the complex $\beta $-plane.
Solving (\ref{5.18}) for $v,$ we get%
\begin{equation}
v(\beta )=\frac{\varphi _{b+1}\beta -e^{i\delta }\Delta \varphi _{b}}{%
-\theta _{b+1}\beta +e^{i\delta }\Delta \theta _{b}}.  \label{5.20}
\end{equation}%
Thus, condition (\ref{5.16}) will be satisf\/ied if we choose $v$ by (\ref%
{5.20}) for pure imaginary values of $\beta .$ On the other hand, when $%
\beta $ runs in (\ref{5.20}) the imaginary axis, $v(\beta )$ describes a
circle $C_{b}$ in the $v$-plane. The center of the circle is symmetric point
of the point at inf\/inity with respect to the circle. Since%
\[
v(\beta ^{\prime })=\infty ,\qquad \text{where} \qquad \beta ^{\prime
}=e^{i\delta }\frac{\Delta \theta _{b}}{\theta _{b+1}},
\]%
the point%
\[
\beta _{0}=-\overline{\beta ^{\prime }}=-e^{-i\delta }\frac{\Delta \overline{%
\theta }_{b}}{\overline{\theta }_{b+1}}
\]%
which is symmetric point of the point $\beta ^{\prime }$ with respect to the
imaginary axis of the $\beta $-plane, is mapped onto the center of $C_{b}$.
So the center of $C_{b}$ is located at the point%
\begin{equation}
v(\beta _{0})=-\frac{e^{-i\delta }\varphi _{b+1}\Delta \overline{\theta }%
_{b}+e^{i\delta }(\Delta \varphi _{b})\overline{\theta }_{b+1}}{e^{-i\delta
}\theta _{b+1}\Delta \overline{\theta }_{b}+e^{i\delta }\overline{\theta }%
_{b+1}\Delta \theta _{b}}.  \label{5.21}
\end{equation}%
Note that the denominator in (\ref{5.21}) is dif\/ferent from zero. This fact
follows from the equality%
\begin{gather}
e^{-i\delta }\theta _{b+1}\Delta \overline{\theta }_{b}+e^{i\delta }%
\overline{\theta }_{b+1}\Delta \theta _{b}=2{\rm Re}\big\{ e^{i\delta
}(\Delta \theta _{n})\overline{\theta }_{n+1}\big|_{1}^{b}\big\}\nonumber\\
\phantom{e^{-i\delta }\theta _{b+1}\Delta \overline{\theta }_{b}+e^{i\delta }%
\overline{\theta }_{b+1}\Delta \theta _{b}}{}
=(2\cos \delta )\sum_{n=2}^{b}\big( \left\vert \Delta \theta
_{n}\right\vert ^{2}+q_{n}\left\vert \theta _{n}\right\vert ^{2}\big),
\label{5.22}
\end{gather}%
which can be derived as (\ref{5.14}) taking into account (\ref{5.10}). The
radius $R_{b}$ of the circle $C_{b}$ is equal to the distance between the
center $v(\beta _{0})$ of $C_{b}$ and the point $v(0)$ on the circle.
Calculating the dif\/ference $v(\beta _{0})-v(0)$ by using (\ref{5.19})--(\ref{5.22}) we easily f\/ind that%
\[
R_{b}=\frac{1}{(2\cos \delta )\sum\limits_{n=2}^{b}\big( \left\vert \Delta \theta
_{n}\right\vert ^{2}+q_{n}\left\vert \theta _{n}\right\vert ^{2}\big) }.
\]%
Further, since%
\[
{\rm Re}(e^{i\delta }\overline{\theta }_{b+1}\Delta \theta
_{b})=-\left\vert \theta _{b+1}\right\vert ^{2}{\rm Re}\,\beta _{0},
\]%
we get from (\ref{5.22})%
\[
(\cos \delta )\sum_{n=2}^{b}\big( \left\vert \Delta \theta _{n}\right\vert
^{2}+q_{n}\left\vert \theta _{n}\right\vert ^{2}\big) =-\left\vert \theta
_{b+1}\right\vert ^{2}{\rm Re}\, \beta _{0}.
\]%
Therefore ${\rm Re}\,\beta _{0}<0.$ This means that the left half-plane of
the $\beta $-plane is mapped onto the interior of the circle $C_{b}.$
Consequently, $v(\beta )$ belongs to the interior of the circle $C_{b}$ if
and only if ${\rm Re}\,\beta <0.$ This inequality is equivalent by (\ref{5.14}), (\ref{5.18}) to%
\begin{equation}
(\cos \delta )\sum_{n=2}^{b}\big( \left\vert \Delta \psi _{n}\right\vert
^{2}+q_{n}\left\vert \psi _{n}\right\vert ^{2}\big) <{\rm Re}\big(ve^{-i\delta }\big).  \label{5.23}
\end{equation}

Thus, $v$ belongs to the interior of the circle $C_{b}$ if and only if the
inequality (\ref{5.23}) holds and~$v$ lies on the circle $C_{b}$ if and only
if%
\[
(\cos \delta )\sum_{n=2}^{b}\big( \left\vert \Delta \psi _{n}\right\vert
^{2}+q_{n}\left\vert \psi _{n}\right\vert ^{2}\big) ={\rm Re}
\big(ve^{-i\delta }\big).
\]

Now let $b_{2}>b_{1}.$ Then if $v$ is inside or on $C_{b_{2}}$%
\[
(\cos \delta )\sum_{n=2}^{b_{1}}\big( \left\vert \Delta \psi
_{n}\right\vert ^{2}+q_{n}\left\vert \psi _{n}\right\vert ^{2}\big) <(\cos
\delta )\sum_{n=2}^{b_{2}}\big( \left\vert \Delta \psi _{n}\right\vert
^{2}+q_{n}\left\vert \psi _{n}\right\vert ^{2}\big) \leq {\rm Re}
\big(ve^{-i\delta }\big)
\]%
and therefore $v$ is inside $C_{b_{1}}.$ This means $C_{b_{1}}$ contains $%
C_{b_{2}}$ in its interior if $b_{2}>b_{1}.$ It follows that, as $%
b\rightarrow \infty ,$ the circles $C_{b}$ converge either to a limit-circle
or to a limit-point. If $\widehat{v}$ is the limit-point or any point on the
limit-circle, then $\widehat{v}$ is inside any $C_{b}.$ Hence%
\[
(\cos \delta )\sum_{n=2}^{b}\big( \left\vert \Delta \psi _{n}\right\vert
^{2}+q_{n}\left\vert \psi _{n}\right\vert ^{2}\big) <{\rm Re}\big(\widehat{v}%
e^{-i\delta }\big),
\]%
where%
\begin{equation}
\psi _{n}=\varphi _{n}+\widehat{v}\theta _{n},\qquad n\in
\mathbb{Z}
,  \label{5.24}
\end{equation}%
and letting $b\rightarrow \infty $ we get%
\begin{equation}
(\cos \delta )\sum_{n=2}^{\infty }\big( \left\vert \Delta \psi
_{n}\right\vert ^{2}+q_{n}\left\vert \psi _{n}\right\vert ^{2}\big) \leq
{\rm Re}\big(\widehat{v}e^{-i\delta }\big).  \label{5.25}
\end{equation}%
It also follows that%
\begin{equation}
{\rm Re}\big(\widehat{v}e^{-i\delta }\big)>0.  \label{5.26}
\end{equation}%
Since $q_{n}\geq c>0,$ (\ref{5.25}) implies that for the solution $\psi
=(\psi _{n})$ def\/ined by (\ref{5.24}) we have (\ref{5.4}). Thus, the
statement of the theorem concerning the solution $\psi =(\psi _{n})$ is
proved.

Let us now show existence of the solution $\chi =(\chi _{n})$ satisfying (%
\ref{5.4}). We seek the desired solution $\theta =(\theta _{n})$ of problem (%
\ref{5.1}), (\ref{5.2}) in the form%
\[
\chi _{n}=\varphi _{n}+u\theta _{n},\qquad n\in
\mathbb{Z}
,
\]
where $u$ is a complex constant to be determined.

Take an arbitrary integer $a\leq -1.$ Applying (\ref{5.8}) to the equations%
\begin{gather*}
-\Delta ^{2}\chi _{n-1}+q_{n}\chi _{n}=0,\qquad n\in
\mathbb{Z}
_{0},
\\
\chi _{-1}=\chi _{1},\qquad \Delta \chi _{-1}=e^{2i\delta }\Delta \chi
_{1},
\end{gather*}
we get%
\[
\sum_{n=a}^{-1}\big( \left\vert \Delta \chi _{n}\right\vert
^{2}+q_{n}\left\vert \chi _{n}\right\vert ^{2}\big) =(\Delta \chi _{n})%
\overline{\chi }_{n+1}\big|_{a-1}^{-1}.
\]%
Since%
\[
\Delta \chi _{-1}=-e^{2i\delta },\qquad \chi _{0}=1-e^{2i\delta }+u,
\]%
we have%
\begin{equation}
\sum_{n=a}^{-1}\big( \left\vert \Delta \chi _{n}\right\vert
^{2}+q_{n}\left\vert \chi _{n}\right\vert ^{2}\big) =-e^{2i\delta }+1-%
\overline{u}e^{2i\delta }-(\Delta \chi _{a-1})\overline{\chi }_{a}.
\label{5.27}
\end{equation}%
Multiply both sides of (\ref{5.27}) by $e^{-i\delta }$ and take then the
real part of both sides to get%
\begin{equation}
(\cos \delta )\sum_{n=a}^{-1}\big( \left\vert \Delta \chi _{n}\right\vert
^{2}+q_{n}\left\vert \chi _{n}\right\vert ^{2}\big) =-{\rm Re}
\big(ue^{-i\delta }\big)-{\rm Re}\big\{ e^{-i\delta }(\Delta \chi _{a-1})%
\overline{\chi }_{a}\big\} .  \label{5.28}
\end{equation}

Now we choose $u$ so that to have%
\begin{equation}
{\rm Re}\big\{e^{-i\delta }(\Delta \chi _{a-1})\overline{\chi }_{a}\big\}=0.
\label{5.29}
\end{equation}%
Since%
\[
{\rm Re}\big\{e^{-i\delta }(\Delta \chi _{a-1})\overline{\chi }%
_{a}\big\}=\left\vert \chi _{a}\right\vert ^{2}{\rm Re}\left\{ e^{-i\delta }%
\frac{\Delta \chi _{a-1}}{\chi _{a}}\right\} ,
\]%
it is suf\/f\/icient for (\ref{5.29}) to have%
\begin{equation}
{\rm Re} \left\{ e^{-i\delta }\frac{\Delta \chi _{a-1}}{\chi _{a}}\right\}
=0.  \label{5.30}
\end{equation}%
Then (\ref{5.28}) becomes
\begin{equation*}
(\cos \delta )\sum_{n=a}^{-1}\left( \left\vert \Delta \chi _{n}\right\vert
^{2}+q_{n}\left\vert \chi _{n}\right\vert ^{2}\right) =-{\rm Re}\big(ue^{-i\delta }\big).  %\label{5.31}
\end{equation*}%
The condition (\ref{5.30}) can be written as%
\begin{equation}
e^{-i\delta }\frac{\Delta \varphi _{a-1}+u\Delta \theta _{a-1}}{\varphi
_{a}+u\theta _{a}}=\alpha ,  \label{5.32}
\end{equation}%
where $\alpha $ is a pure imaginary number ($\alpha =it$, $t\in
\mathbb{R}).$ Note that%
\begin{equation}
\varphi _{a}\Delta \theta _{a-1}-(\Delta \varphi _{a-1})\theta
_{a}=W_{a-1}(\varphi ,\theta )=e^{2i\delta }\neq 0  \label{5.33}
\end{equation}%
by (\ref{5.11}). Therefore (\ref{5.32}) def\/ines a linear-fractional
transformation of the complex $u$-plane onto the complex $\alpha $-plane.
Solving (\ref{5.32}) for $u,$ we get%
\begin{equation}
u(\alpha )=\frac{\varphi _{a}\alpha -e^{-i\delta }\Delta \varphi _{a-1}}{%
-\theta _{a}\alpha +e^{-i\delta }\Delta \theta _{a-1}}.  \label{5.34}
\end{equation}%
Thus, condition (\ref{5.30}) will be satisf\/ied if we choose $u$ by (\ref%
{5.34}) for pure imaginary values of $\alpha .$ On the other hand, when $%
\alpha $ runs in (\ref{5.34}) the imaginary axis, $u(\alpha )$ describes a
circle $K_{a}$ in the $u$-plane. The center of the circle is symmetric point
of the point at inf\/inity with respect to the circle. Since%
\[
u(\alpha ^{\prime })=\infty ,\qquad \text{where} \qquad \alpha ^{\prime
}=e^{-i\delta }\frac{\Delta \theta _{a-1}}{\theta _{a}},
\]%
the point%
\[
\alpha _{0}=-\overline{\alpha ^{\prime }}=-e^{i\delta }\frac{\Delta
\overline{\theta }_{a-1}}{\overline{\theta }_{a}}
\]%
which is symmetric point of the point $\alpha ^{\prime }$ with respect to
the imaginary axis of the $\alpha $-plane, is mapped onto the center of $%
K_{a}.$ So the center of $K_{a}$ is located at the point%
\begin{equation}
u(\alpha _{0})=-\frac{e^{i\delta }\varphi _{a}\Delta \overline{\theta }%
_{a-1}+e^{-i\delta }(\Delta \varphi _{a-1})\overline{\theta }_{a}}{%
e^{i\delta }\theta _{a}\Delta \overline{\theta }_{a-1}+e^{-i\delta }%
\overline{\theta }_{a}\Delta \theta _{a-1}}.  \label{5.35}
\end{equation}%
Note that the denominator in (\ref{5.35}) is dif\/ferent from zero. This fact
follows from the equality%
\begin{gather}
e^{-i\delta }\theta _{a}\Delta \overline{\theta }_{a-1}+e^{i\delta }%
\overline{\theta }_{a}\Delta \theta _{a-1}=2{\rm Re}\big\{ e^{i\delta
}(\Delta \theta _{a-1})\overline{\theta }_{a}\big\}\nonumber\\
\qquad {}=-(2\cos \delta )\sum_{n=a}^{-1}\big( \left\vert \Delta \theta
_{n}\right\vert ^{2}+q_{n}\left\vert \theta _{n}\right\vert ^{2}\big)
\label{5.36}
\end{gather}%
which can be derived as (\ref{5.28}) taking into account $\Delta \theta
_{-1}=0,$ $\theta _{0}=1.$ Calculating the dif\/ference $u(\alpha _{0})-u(0)$
we easily f\/ind the radius $R_{a}=\left\vert u(\alpha _{0})-u(0)\right\vert $
of the circle $K_{a},$ using (\ref{5.33})--(\ref{5.36}),%
\[
R_{a}=\frac{1}{(2\cos \delta )\sum\limits_{n=a}^{-1}\big( \left\vert \Delta \theta
_{n}\right\vert ^{2}+q_{n}\left\vert \theta _{n}\right\vert ^{2}\big) }.
\]%
Further, since%
\[
{\rm Re}\big(e^{-i\delta }\overline{\theta }_{a}\Delta \theta
_{a-1}\big)=-\left\vert \theta _{a}\right\vert ^{2}{\rm Re}\,\alpha _{0},
\]%
we get from (\ref{5.36})%
\[
(\cos \delta )\sum_{n=a}^{-1}\big( \left\vert \Delta \theta _{n}\right\vert
^{2}+q_{n}\left\vert \theta _{n}\right\vert ^{2}\big) =\left\vert \theta
_{a}\right\vert ^{2}{\rm Re}\,\alpha _{0}.
\]%
Therefore ${\rm Re}\,\alpha _{0}>0.$ This means that the right half-plane
of the $\alpha $-plane is mapped onto the interior of the circle $K_{a}.$
Consequently, $u(\alpha )$ lies inside the circle $K_{a}$ if and only if $%
{\rm Re}\,\alpha >0.$ This inequality is equivalent by (\ref{5.28}), (\ref{5.32}) to
\begin{equation}
(\cos \delta )\sum_{n=a}^{-1}\big( \left\vert \Delta \chi _{n}\right\vert
^{2}+q_{n}\left\vert \chi _{n}\right\vert ^{2}\big) <-{\rm Re}
\big(ue^{-i\delta }\big).  \label{5.37}
\end{equation}

Thus, $u$ lies inside the circle $K_{a}$ if and only if the inequality (\ref%
{5.37}) holds and $u$ lies on the circle $K_{a}$ if and only if%
\[
(\cos \delta )\sum_{n=a}^{-1}\big( \left\vert \Delta \chi _{n}\right\vert
^{2}+q_{n}\left\vert \chi _{n}\right\vert ^{2}\big) =-{\rm Re}
\big(ue^{-i\delta }\big).
\]

Now let $a_{2}<a_{1}.$ Then if $u$ is inside or on $K_{a_{2}}$%
\[
(\cos \delta )\sum_{n=a_{1}}^{-1}\big( \left\vert \Delta \chi
_{n}\right\vert ^{2}+q_{n}\left\vert \chi _{n}\right\vert ^{2}\big) <(\cos
\delta )\sum_{n=a_{2}}^{-1}\big( \left\vert \Delta \chi _{n}\right\vert
^{2}+q_{n}\left\vert \chi _{n}\right\vert ^{2}\big) \leq -{\rm Re}
\big(ue^{-i\delta }\big)
\]%
and therefore $u$ is inside $K_{a_{1}}$. This means $K_{a_{1}}$ contains $%
K_{a_{2}}$ in its interior if $a_{2}<a_{1}.$ It follows that, as $%
a\rightarrow -\infty ,$ the circles $K_{a}$ converge either to a
limit-circle or to a limit-point. If $\widehat{u}$ is the limit-point or any
point on the limit-circle, then $\widehat{u}$ is inside any $K_{a}.$ Hence%
\[
(\cos \delta )\sum_{n=a}^{-1}\big( \left\vert \Delta \chi _{n}\right\vert
^{2}+q_{n}\left\vert \chi _{n}\right\vert ^{2}\big) <-{\rm Re}\big(\widehat{u}%
e^{-i\delta }\big),
\]%
where%
\begin{equation}
\chi _{n}=\varphi _{n}+\widehat{u}\theta _{n},\qquad n\in
\mathbb{Z}
,  \label{5.38}
\end{equation}%
and letting $a\rightarrow -\infty $ we get%
\begin{equation}
(\cos \delta )\sum_{-\infty }^{n=-1}\big( \left\vert \Delta \chi
_{n}\right\vert ^{2}+q_{n}\left\vert \chi _{n}\right\vert ^{2}\big) \leq -%
{\rm Re}\big(\widehat{u}e^{-i\delta }\big).  \label{5.39}
\end{equation}%
It also follows that%
\begin{equation}
{\rm Re}\big(\widehat{u}e^{-i\delta }\big)<0.  \label{5.40}
\end{equation}%
Since $q_{n}\geq c>0,$ (\ref{5.39}) implies that for the solution $\chi
=(\chi _{n})$ def\/ined by (\ref{5.38}) we have (\ref{5.4}). Thus, the
statement of the theorem concerning the solution $\chi =(\chi _{n})$ is also
proved.

Finally, let us show that the solutions $\psi =(\psi _{n})$ and $\chi =(\chi
_{n})$ def\/ined by (\ref{5.24}) and (\ref{5.38}), respectively, are linearly
independent. We have%
\begin{equation}
W_{n}(\psi ,\chi )=W_{n}(\varphi +\widehat{v}\theta ,\varphi +\widehat{u}%
\theta )=(\widehat{u}-\widehat{v})W_{n}(\varphi ,\theta ).  \label{5.41}
\end{equation}%
Next, $W_{n}(\varphi ,\theta )\neq 0$ by (\ref{5.11}) and $\widehat{u}\neq
\widehat{v}$ by (\ref{5.26}), (\ref{5.40}). Therefore $W_{n}(\psi ,\chi
)\neq 0$ and hence $\psi $ and $\chi $ are linearly independent by Theorem %
\ref{Th4.5}.
\end{proof}

\section[The inverse operator $L^{-1}$]{The inverse operator $\boldsymbol{L^{-1}}$}\label{section6}

The following lemma will play crucial role in this and next sections.

\begin{lemma}
\label{Lem6.1} Let us set%
\begin{equation}
\sigma _{n}=\left\{
\begin{array}{lll}
e^{-i\delta } & \text{for} & n\leq -1, \\
e^{i\delta } & \text{for} & n\geq 0.%
\end{array}%
\right.  \label{6.1}
\end{equation}%
Under the conditions of Lemma {\rm \ref{Lem5.1}} the following formula holds:%
\begin{gather}
(\cos \delta )\left( \sum_{n=a}^{-1}+\sum_{n=2}^{b}\right) \big(
\left\vert \Delta y_{n}\right\vert ^{2}+q_{n}\left\vert y_{n}\right\vert
^{2}\big)  \nonumber \\
\qquad ={\rm Re}\left\{ \sigma _{n}(\Delta y_{n})\overline{y}_{n+1}\big|
_{a-1}^{b}+\left( \sum_{n=a}^{-1}+\sum_{n=2}^{b}\right) \sigma _{n}f_{n}%
\overline{y}_{n}\right\} .  \label{6.2}
\end{gather}
\end{lemma}

\begin{proof}
To prove (\ref{6.2}) we multiply (\ref{5.8}) by $e^{-i\delta }$ and (\ref%
{5.7}) by $e^{i\delta }$ and add together to get%
\begin{gather}
\left( e^{-i\delta }\sum_{n=a}^{-1}+e^{i\delta }\sum_{n=2}^{b}\right) \big(
\left\vert \Delta y_{n}\right\vert ^{2}+q_{n}\left\vert y_{n}\right\vert
^{2}\big)
\nonumber\\
\qquad =e^{-i\delta }(\Delta y_{n})\overline{y}_{n+1}\big|_{a-1}^{-1}+e^{i\delta
}(\Delta y_{n})\overline{y}_{n+1}\big|_{1}^{b}+\left( e^{-i\delta
}\sum_{n=a}^{-1}+e^{i\delta }\sum_{n=2}^{b}\right) f_{n}\overline{y}_{n}.
\label{6.3}
\end{gather}%
Next, using the conditions (\ref{5.6}) we have%
\begin{gather*}
e^{-i\delta }(\Delta y_{-1})\overline{y}_{0}-e^{i\delta }(\Delta y_{1})%
\overline{y}_{2}=e^{i\delta }(\Delta y_{1})\overline{y}_{0}-e^{i\delta
}(\Delta y_{1})\overline{y}_{2}\\
\qquad{}
=e^{i\delta }(\Delta y_{1})(\overline{y}_{0}-\overline{y}_{2})=e^{i\delta
}(\Delta y_{1})(\overline{y}_{0}-\overline{y}_{1}+\overline{y}_{1}-\overline{%
y}_{2})
\\
\qquad{}=e^{i\delta }(\Delta y_{1})(\overline{y}_{0}-\overline{y}_{-1}+\overline{y}%
_{1}-\overline{y}_{2})=e^{i\delta }(\Delta y_{1})(\overline{\Delta y_{-1}}-%
\overline{\Delta y_{1}})
\\
\qquad{}
=-e^{i\delta }\left\vert \Delta y_{1}\right\vert ^{2}+e^{i\delta }(\Delta
y_{1})e^{-2i\delta }\overline{\Delta y_{1}}=-e^{i\delta }\left\vert \Delta
y_{1}\right\vert ^{2}+e^{-i\delta }\left\vert \Delta y_{1}\right\vert ^{2}
=-2i(\sin \delta )\left\vert \Delta y_{1}\right\vert ^{2}.
\end{gather*}
Therefore taking (\ref{6.1}) into account we can rewrite (\ref{6.3}) in the
form%
\begin{gather}
\left( \sum_{n=a}^{-1}+\sum_{n=2}^{b}\right) \sigma _{n}\big( \left\vert
\Delta y_{n}\right\vert ^{2}+q_{n}\left\vert y_{n}\right\vert ^{2}\big)
\nonumber\\
\qquad{} =-2i(\sin \delta )\left\vert \Delta y_{1}\right\vert ^{2}+\sigma _{n}(\Delta
y_{n})\overline{y}_{n+1}\big|_{a-1}^{b}+\left(
\sum_{n=a}^{-1}+\sum_{n=2}^{b}\right) \sigma _{n}f_{n}\overline{y}_{n}.
\label{6.4}
\end{gather}%
Taking in (\ref{6.4}) the real parts of both sides and taking into account
that ${\rm Re}\,\sigma _{n}=\cos \delta $ for all $n$, we obtain (\ref{6.2}).
\end{proof}

Let $L:D\subset l_{0}^{2}\rightarrow l_{0}^{2}$ be the operator def\/ined
above in Section~\ref{section2}. Further, let $\psi =(\psi _{n})$ and $\chi
=(\chi _{n}),$ where $n\in
\mathbb{Z}
$, be solutions of problem (\ref{5.1}), (\ref{5.2}), constructed in Theorem~\ref{Th5.2}. Let us introduce the discrete Green function%
\begin{equation*}
G_{nk}=\frac{1}{W_{k}(\psi ,\chi )}\left\{
\begin{array}{lll}
\chi _{k}\psi _{n} & \text{if} & k\leq n, \\
\chi _{n}\psi _{k} & \text{if} & k\geq n,%
\end{array}%
\right. % \label{6.5}
\end{equation*}%
of discrete variables $k,n\in
\mathbb{Z}
.$ Note that by (\ref{5.41}) and (\ref{5.11}), we have%
\begin{equation}
W_{0}(\psi ,\chi )=-(\widehat{u}-\widehat{v})e^{2i\delta },\qquad
W_{k}(\psi ,\chi )=\left\{
\begin{array}{lll}
(\widehat{u}-\widehat{v})e^{2i\delta } & \text{if} & k\leq -1, \\
\widehat{u}-\widehat{v} & \text{if} & k\geq 1,%
\end{array}%
\right.  \label{6.6}
\end{equation}%
and, besides,%
\begin{equation*}
\widehat{u}\neq \widehat{v}  %\label{6.7}
\end{equation*}%
by (\ref{5.26}) and (\ref{5.40}).

\begin{theorem}
\label{Th6.1}Under the condition \eqref{5.3} the inverse operator $L^{-1}$
exists and is a bounded operator defined on the whole space $l_{0}^{2}.$
Next, for every $f=(f_{n})\in l_{0}^{2}$%
\begin{equation}
\big(L^{-1}f\big)_{n}=\sum_{k\in
\mathbb{Z}
_{0}}G_{nk}f_{k},\qquad n\in \mathbb{Z}
_{0},  \label{6.8}
\end{equation}%
and%
\begin{equation}
\big\Vert L^{-1}f\big\Vert \leq \frac{1}{c\cos \delta }\left\Vert
f\right\Vert \qquad \text{for all} \ \ f\in l_{0}^{2},  \label{6.9}
\end{equation}%
where $c$ is a constant from condition \eqref{5.3} and $\delta $ is from \eqref{5.2}, $\left\Vert \cdot \right\Vert $ denotes the norm of space $l_{0}^{2}.$
\end{theorem}

\begin{proof}
Let us show that
\[
\ker L=\left\{ y\in D:Ly=0\right\}
\]
consists only of the zero element. Indeed, if $y\in D$ and $Ly=0,$ then $%
(y)_{n\in
\mathbb{Z}
_{0}}$ satisf\/ies the equation%
\begin{equation}
-\Delta ^{2}y_{n-1}+q_{n}y_{n}=0,\qquad n\in
\mathbb{Z}
_{0},  \label{6.10}
\end{equation}%
in which $y_{0}$ and $y_{1}$ (these values arise in (\ref{6.10}) for $n=-1$
and $n=2,$ respectively) are def\/ined from the equations%
\begin{equation}
y_{-1}=y_{1},\qquad \Delta y_{-1}=e^{2i\delta }\Delta y_{1}.
\label{6.11}
\end{equation}%
Since $\chi $ and $\psi $ form a fundamental system of solutions of (\ref%
{6.10}), (\ref{6.11}), we can write
\[
y_{n}=C_{1}\psi _{n}+C_{2}\chi _{n},\qquad n\in
\mathbb{Z}
,
\]%
with some constants $C_{1}$ and $C_{2}.$ Hence%
\begin{equation}
W_{n}(y,\psi )=C_{1}W_{n}(\psi ,\psi )+C_{2}W_{n}(\chi ,\psi ),\qquad
n\in
\mathbb{Z}
.  \label{6.12}
\end{equation}%
Next, since $y\in l_{0}^{2},$ we have $y_{n}\rightarrow 0$ as $\left\vert
n\right\vert \rightarrow \infty $ and by (\ref{5.4}) we have $\psi
_{n}\rightarrow 0$ as $n\rightarrow \infty $ \ Hence $W_{n}(y,\psi
)\rightarrow 0$ as $n\rightarrow \infty .$ Besides $W_{n}(\psi ,\psi )=0$
for all $n$ and $W_{n}(\chi ,\psi )$ is equal to a nonzero constant for $%
n\geq 1$ by (\ref{6.6}). Therefore taking the limit in (\ref{6.12}) as $%
n\rightarrow \infty $ we get that $C_{2}=0.$ It can similarly be shown, by
considering $W_{n}(y,\chi )$, that $C_{1}=0.$ Thus $y=0.$

It follows that the inverse operator $L^{-1}$ exists. Now take an arbitrary $%
f=(f_{n})_{n\in
\mathbb{Z}
_{0}}\in l_{0}^{2}$ and extend the sequence $(f_{n})_{n\in
\mathbb{Z}
_{0}}$ to the values $n=0$ and $n=1$ by setting%
\begin{equation*}
f_{0}=f_{1}=0.  %\label{6.13}
\end{equation*}%
Let us put%
\begin{gather}
g_{n}=\sum_{k\in
\mathbb{Z}
_{0}}G_{nk}f_{k}=\sum_{k\in
\mathbb{Z}
}G_{nk}f_{k}
=\psi _{n}\sum_{-\infty }^{k=n}\frac{\chi _{k}f_{k}}{W_{k}(\psi ,\chi )}%
+\chi _{n}\sum_{k=n+1}^{\infty }\frac{\psi _{k}f_{k}}{W_{k}(\psi ,\chi )},\qquad
n\in
\mathbb{Z}
.  \label{6.14}
\end{gather}%
Then it is easy to check that this sequence $(g_{n}),$ where $n\in
\mathbb{Z}
$, satisf\/ies the equations%
\begin{gather}
-\Delta ^{2}g_{n-1}+q_{n}g_{n}=f_{n},\qquad n\in
\mathbb{Z}
_{0},  \label{6.15}
\\
g_{-1}=g_{1},\qquad \Delta g_{-1}=e^{2i\delta }\Delta g_{1}.
\label{6.16}
\end{gather}%
We want to show that $g=(g_{n})_{n\in
\mathbb{Z}
_{0}}\in l_{0}^{2}$ and that%
\begin{equation}
\left\Vert g\right\Vert \leq \frac{1}{c\cos \delta }\left\Vert f\right\Vert .
\label{6.17}
\end{equation}%
For this purpose we take the sequences of integers $a_{m}$ and $b_{m}$
def\/ined for any positive integer~$m$, such that%
\[
a_{m}<0<b_{m}\qquad \text{and} \qquad a_{m}\rightarrow -\infty ,\qquad
b_{m}\rightarrow \infty \qquad \text{as} \qquad m\rightarrow \infty .
\]%
Next, for each $m$ we def\/ine the sequence $(f_{n}^{(m)})_{n\in
\mathbb{Z}
}$ by%
\begin{gather}
f_{n}^{(m)}=f_{n}\qquad \text{if} \qquad a_{m}\leq n\leq b_{m},  \label{6.18}
\\
f_{n}^{(m)}=0 \qquad \text{if} \qquad n<a_{m} \qquad \text{or} \qquad n>b_{m},  \label{6.19}
\end{gather}%
and put%
\begin{gather*}
g_{n}^{(m)}=\sum_{k\in
\mathbb{Z}
_{0}}G_{nk}f_{k}^{(m)}=\sum_{k\in
\mathbb{Z}
}G_{nk}f_{k}^{(m)}
=\psi _{n}\sum_{-\infty }^{k=n}\frac{\chi _{k}f_{k}^{(m)}}{W_{k}(\psi ,\chi )%
}+\chi _{n}\sum_{k=n+1}^{\infty }\frac{\psi _{k}f_{k}^{(m)}}{W_{k}(\psi
,\chi )},\qquad n\in \mathbb{Z}
.  %\label{6.20}
\end{gather*}%
It follows that%
\begin{gather}
g_{n}^{(m)}=\left\{
\begin{array}{lll}
\displaystyle \chi _{n}\sum_{k=a_{m}}^{b_{m}}\frac{\psi _{k}f_{k}}{W_{k}(\psi ,\chi )} &
\text{if} & n<a_{m}, \vspace{1mm}\\
\displaystyle \psi _{n}\sum_{k=a_{m}}^{b_{m}}\frac{\chi _{k}f_{k}}{W_{k}(\psi ,\chi )} &
\text{if} & n>b_{m}.%
\end{array}%
\right.  \label{6.21}
\end{gather}%
We have also that, for each $m,$%
\begin{gather}
-\Delta ^{2}g_{n-1}^{(m)}+q_{n}g_{n}^{(m)}=f_{n}^{(m)},\qquad n\in
\mathbb{Z}
_{0},  \label{6.22}
\\
g_{-1}^{(m)}=g_{1}^{(m)},\qquad \Delta g_{-1}^{(m)}=e^{2i\delta }\Delta
g_{1}^{(m)}.  \label{6.23}
\end{gather}%
Fix $m$ and take a positive integer $N$ such that%
\[
-N<a_{m}\qquad \text{and} \qquad b_{m}<N.
\]%
Then applying Lemma \ref{Lem6.1} to (\ref{6.22}), (\ref{6.23}) we can write%
\begin{gather}
(\cos \delta )\left( \sum_{n=-N}^{-1}+\sum_{n=2}^{N}\right) \big(
\big\vert \Delta g_{n}^{(m)}\big\vert ^{2}+q_{n}\big\vert
g_{n}^{(m)}\big\vert ^{2}\big)  \nonumber \\
\qquad {}= {\rm Re}\left\{ \sigma _{n}(\Delta g_{n}^{(m)})\overline{g}%
_{n+1}^{(m)}\big| _{-N-1}^{N}+\left( \sum_{n=-N}^{-1}+\sum_{n=2}^{N}\right)
\sigma _{n}f_{n}^{(m)}\overline{g}_{n}^{(m)}\right\} .  \label{6.24}
\end{gather}%
It follows from (\ref{6.21}) by (\ref{5.4}) that%
\[
\sum_{n\in
\mathbb{Z}
}\big\vert g_{n}^{(m)}\big\vert ^{2}<\infty.
\]
Therefore the sums on the right-hand side of (\ref{6.24}) are convergent as $%
N\rightarrow \infty $ and besides%
\[
{\rm Re}\big\{ \sigma _{n}(\Delta g_{n}^{(m)})\overline{g}_{n+1}^{(m)}\big|
_{-N-1}^{N}\big\} \rightarrow 0\qquad \text{as} \ \  N\rightarrow \infty .
\]%
(note that $\left\vert \sigma _{n}\right\vert =1$ for all $n$ by (\ref{6.1}%
)). Now taking the limit in (\ref{6.24}) as $N\rightarrow \infty ,$ we get%
\begin{equation}
(\cos \delta )\sum_{n\in
\mathbb{Z}
_{0}}\big( \big\vert \Delta g_{n}^{(m)}\big\vert ^{2}+q_{n}\big\vert
g_{n}^{(m)}\big\vert ^{2}\big) ={\rm Re}\sum_{n\in
\mathbb{Z}
_{0}}\sigma _{n}f_{n}^{(m)}\overline{g}_{n}^{(m)}.  \label{6.25}
\end{equation}%
Using the condition (\ref{5.3}) we get from (\ref{6.25}) that%
\begin{gather*}
\sum_{n\in
\mathbb{Z}
_{0}}\big\vert g_{n}^{(m)}\big\vert ^{2}\leq \frac{1}{c\cos \delta }%
{\rm Re}\sum_{n\in
\mathbb{Z}
_{0}}\sigma _{n}f_{n}^{(m)}\overline{g}_{n}^{(m)}
\leq \frac{1}{c\cos \delta }\left\vert \sum_{n\in
\mathbb{Z}
_{0}}\sigma _{n}f_{n}^{(m)}\overline{g}_{n}^{(m)}\right\vert \\
\qquad {} \leq \frac{1}{%
c\cos \delta }\sum_{n\in
\mathbb{Z}
_{0}}\big\vert f_{n}^{(m)}\overline{g}_{n}^{(m)}\big\vert
\leq \frac{1}{c\cos \delta }\left\{ \sum_{n\in
\mathbb{Z}
_{0}}\big\vert f_{n}^{(m)}\big\vert ^{2}\right\} ^{1/2}\left\{ \sum_{n\in
\mathbb{Z}
_{0}}\big\vert g_{n}^{(m)}\big\vert ^{2}\right\} ^{1/2}.
\end{gather*}
Hence%
\[
\left\{ \sum_{n\in
\mathbb{Z}
_{0}}\big\vert g_{n}^{(m)}\big\vert ^{2}\right\} ^{1/2}\leq \frac{1}{%
c\cos \delta }\left\{ \sum_{n\in
\mathbb{Z}
_{0}}\big\vert f_{n}^{(m)}\big\vert ^{2}\right\} ^{1/2},
\]%
that is,%
\begin{equation}
\big\Vert g^{(m)}\big\Vert \leq \frac{1}{c\cos \delta }\big\Vert
f^{(m)}\big\Vert .  \label{6.26}
\end{equation}

Writing (\ref{6.22}), (\ref{6.23}) for $m=m_{1}$ and $m=m_{2},$ and
subtracting the obtained equations side-by-side, we get%
\begin{gather*}
-\Delta
^{2}\big[g_{n-1}^{(m_{1})}-g_{n-1}^{(m_{2})}\big]+q_{n}\big[g_{n}^{(m_{1})}-g_{n}^{(m_{2})}\big]=f_{n}^{(m_{1})}-f_{n}^{(m_{2})},%
\qquad n\in
\mathbb{Z}
_{0},
\\
g_{-1}^{(m_{1})}-g_{-1}^{(m_{2})}=g_{1}^{(m_{1})}-g_{1}^{(m_{2})},\qquad \Delta \big[ g_{-1}^{(m_{1})}-g_{-1}^{(m_{2})}\big]=e^{2i\delta }\Delta
\big[ g_{1}^{(m_{1})}-g_{1}^{(m_{2})}\big].
\end{gather*}
Hence, repeating the same reasonings as above, we get%
\[
\big\Vert g^{(m_{1})}-g^{(m_{2})}\big\Vert \leq \frac{1}{c\cos \delta }%
\big\Vert f^{(m_{1})}-f^{(m_{2})}\big\Vert .
\]%
It follows that $g^{(m)}$ converges in $l_{0}^{2}$ to an element $\widetilde{%
g}$ as $m\rightarrow \infty .$ On the other hand, it can be seen from (\ref{6.14}), (\ref{6.21}) taking into account (\ref{6.18}), (\ref{6.19}) that%
\[
g_{n}^{(m)}\rightarrow g_{n}\qquad \text{as} \ \ m\rightarrow \infty ,
\]%
for each $n$. Consequently, $\widetilde{g}=g$ and hence $g\in l_{0}^{2}.$
Passing in (\ref{6.26}) to the limit as $m\rightarrow \infty ,$ we get (\ref{6.17}).

Next, from (\ref{6.15}) we have%
\[
q_{n}g_{n}=f_{n}+g_{n-1}-2g_{n}+g_{n+1},\qquad n\in
\mathbb{Z}
_{0}.
\]
Hence $(q_{n}g_{n})_{n\in
\mathbb{Z}
_{0}}\in l_{0}^{2}.$ Therefore $g\in D,$ where $D$ is the domain of the
operator~$L$. If we def\/ine an operator $B:l_{0}^{2}\rightarrow l_{0}^{2}$ by
the formula $Bf=g,$ where $f=(f_{n})_{n\in
\mathbb{Z}
_{0}}\in l_{0}^{2}$ and $g=(g_{n})_{n\in
\mathbb{Z}
_{0}}$ with~$g_{n}$ def\/ined by (\ref{6.14}), then we get by (\ref{6.15}), (\ref{6.16}) that $LBf=f.$ Therefore $B$ is the inverse of the operator $%
L:B=L^{-1},$ so that $g=L^{-1}f$ and from (\ref{6.14}) and (\ref{6.17}) we
get~(\ref{6.8}) and~(\ref{6.9}), respectively.
\end{proof}

\section[Completely continuity of the operator $L^{-1}$]{Completely continuity of the operator $\boldsymbol{L^{-1}}$}\label{section7}

In this section we will show that the operator $L^{-1}$ is completely
continuous, that is, it is continuous and maps bounded sets into relatively
compact sets.

\begin{theorem}
\label{Th7.1}Let%
\begin{equation}
q_{n}\geq c>0\qquad \text{for} \ \ n\in
\mathbb{Z}_{0},  \label{7.1}
\end{equation}%
and
\begin{equation}
\lim_{\left\vert n\right\vert \rightarrow \infty }q_{n}=\infty .  \label{7.2}
\end{equation}%
Then the operator $L^{-1}$ is completely continuous.
\end{theorem}

\begin{proof}
The operator $L^{-1}$ is continuous in virtue of (\ref{6.9}) that holds
under the condition (\ref{7.1}). In order to show that $L^{-1}$ maps bounded
sets into relatively compact sets consider any bounded set $X$ in $l_{0}^{2}$,
\[
X=\big\{ f\in l_{0}^{2}:\left\Vert f\right\Vert \leq d\big\} ,
\]%
and prove that $L^{-1}(X)=Y$ is relatively compact in $l_{0}^{2}.$ To this
end, we use the following known (see~\cite{[17]}) criterion for the relative
compactness in $l_{0}^{2}:$ \textit{A set $Y\subset l_{0}^{2}$
is relatively compact if and only if  $Y$ is
bounded and for every $\varepsilon >0$ there exists a positive
integer $n_{0}$ (depending only on~$\varepsilon $) such
that}
\[
\sum_{\left\vert n\right\vert >n_{0}}\left\vert y_{n}\right\vert ^{2}\leq
\varepsilon \qquad \text{\textit{for all}} \ \ y\in Y.
\]

Take an arbitrary $f\in X$ and set
\[
L^{-1}f=y.
\]
Then $Ly=f$ or explicitly%
\begin{equation}
-\Delta ^{2}y_{n-1}+q_{n}y_{n}=f_{n},\qquad n\in
\mathbb{Z}
_{0},  \label{7.3}
\end{equation}%
where $y_{0}$ and $y_{1}$ are def\/ined from the equations%
\begin{equation}
y_{-1}=y_{1},\qquad \Delta y_{-1}=e^{2i\delta }\Delta y_{1}.
\label{7.4}
\end{equation}
Note that $y_{0}$ and $y_{1}$ are needed when we write out equation (\ref{7.3}) for $n=-1$ and $n=2,$ respectively.

Applying Lemma \ref{Lem6.1} to (\ref{7.3}), (\ref{7.4}), we get that for any
integers $a\leq -1$ and $b\geq 2,$%
\begin{gather}
(\cos \delta )\left( \sum_{n=a}^{-1}+\sum_{n=2}^{b}\right) \big(
\left\vert \Delta y_{n}\right\vert ^{2}+q_{n}\left\vert y_{n}\right\vert
^{2}\big)  \nonumber \\
\qquad {} ={\rm Re}\left\{ \sigma _{n}(\Delta y_{n})\overline{y}_{n+1}\big|_{a-1}^{b}+\left( \sum_{n=a}^{-1}+\sum_{n=2}^{b}\right) \sigma _{n}f_{n}%
\overline{y}_{n}\right\} ,  \label{7.5}
\end{gather}
where $\sigma _{n}$ is def\/ined by (\ref{6.1}).

Since $f,y\in l_{0}^{2}$ and $\left\vert \sigma _{n}\right\vert =1,$ the
sums on the right-hand side of (\ref{7.5}) are convergent as $a\rightarrow
-\infty ,$ $b\rightarrow \infty .$ Also from $y\in l_{0}^{2}$ it follows
that $y_{n}\rightarrow 0$ as $\left\vert n\right\vert \rightarrow \infty $
so that%
\[
\sigma _{n}(\Delta y_{n})\overline{y}_{n+1}\big|_{a-1}^{b}\rightarrow 0\qquad \text{as} \qquad a\rightarrow -\infty ,\qquad b\rightarrow \infty .
\]
Consequently, we arrive at the equality%
\[
(\cos \delta )\sum_{n\in
\mathbb{Z}
_{0}}\big( \left\vert \Delta y_{n}\right\vert ^{2}+q_{n}\left\vert
y_{n}\right\vert ^{2}\big) ={\rm Re}\sum_{n\in
\mathbb{Z}
_{0}}\sigma _{n}f_{n}\overline{y}_{n}.
\]%
Hence%
\begin{equation}
(\cos \delta )\sum_{n\in
\mathbb{Z}
_{0}}q_{n}\left\vert y_{n}\right\vert ^{2}\leq {\rm Re}\sum_{n\in
\mathbb{Z}
_{0}}\sigma _{n}f_{n}\overline{y}_{n}  \label{7.6}
\end{equation}%
and therefore using (\ref{7.1}) and $\left\Vert f\right\Vert \leq d,$ we get{\samepage
\begin{equation}
\left\Vert y\right\Vert \leq \frac{d}{c\cos \delta }\qquad \text{for all} \ \
y\in Y.  \label{7.7}
\end{equation}%
This means that the set $Y=L^{-1}(X)$ is bounded.}

From (\ref{7.6}) we also have, using (\ref{7.7}),%
\begin{equation}
\sum_{n\in \mathbf{%
%TCIMACRO{\U{2124} }%
%BeginExpansion
\mathbb{Z}
%EndExpansion
}_{0}}q_{n}\left\vert y_{n}\right\vert ^{2}\leq \frac{d^{2}}{c\cos
^{2}\delta } \qquad \text{for all} \ \ y\in Y.  \label{7.8}
\end{equation}

Take now an arbitrary $\varepsilon >0$. By condition (\ref{7.2}) we can
choose a positive integer $n_{0}$ such that
\[
q_{n}\geq \frac{d^{2}}{\varepsilon c\cos ^{2}\delta } \qquad \text{for} \ \
\left\vert n\right\vert >n_{0}.
\]%
Then we get from (\ref{7.8}) that
\[
\sum_{\left\vert n\right\vert >n_{0}}\left\vert y_{n}\right\vert ^{2}\leq
\varepsilon \qquad \text{for all} \ \ y\in Y.
\]

Thus the completely continuity of the operator $L^{-1}$ is proved.
\end{proof}

\begin{corollary}
\label{Cor7.2} The operator $A=M^{-1}L$ is invertible and its inverse $A^{-1}=L^{-1}M$ is a completely continuous operator to be a product of
completely continuous operator with bounded operator. Therefore the spectrum
of the operator $A$ is discrete.
\end{corollary}

\section{Conclusions}\label{section8}

In this paper we have explored a new class of discrete non-Hermitian quantum
systems. The concept of the spectrum for the considered discrete system is
introduced and discreteness of the spectrum is proved under some simple
conditions.

As a tool for the investigation we have established main statements for
second order linear dif\/ference equations with impulse conditions (transition
conditions). We have chosen a suitable (inf\/inite-dimensional) Hilbert space
and def\/ined the main linear operator $A$ so that the spectrum of the problem
in question coincides with the spectrum of $A.$ Next, we have constructed
the inverse $A^{-1}$ of the operator $A$ by using an appropriate discrete
Green function. Finally, we have shown that the inverse operator $A^{-1}$ is
completely continuous. This implies, in particular, discreteness of the
spectrum of $A.$

\subsection*{Acknowledgements}

This work was supported by Grant 106T549 from the Scientif\/ic and
Technological Research Council of Turkey (TUBITAK). The author thanks Mesude
Saglam and Gusein Guseinov for useful discussions.

\pdfbookmark[1]{References}{ref}
\LastPageEnding

\end{document}